\newif\ifarxiv
\arxivtrue       

\documentclass[aps,prd,physrev,eqsecnum,amsmath,amssymb,nofootinbib,twocolumn]{revtex4-2}

\usepackage{hyperref}
\usepackage{amsmath}
\usepackage{graphicx}
\usepackage{amsthm}
\usepackage{relsize}
\usepackage{aas_macros}
\usepackage{bm}
\usepackage{color}
\usepackage{enumitem}
\usepackage{multirow}
\usepackage{dsfont}
\usepackage{array}[=2016-10-06]
\usepackage{longtable}

\usepackage[normalem]{ulem} 

\ifarxiv

  \usepackage{xcolor}
  \definecolor{Gray}{gray}{0.42}    

\else
\fi

\newcommand{\contcaption}[1]{\addtocounter{table}{-1}\caption{(Continued). #1}}
\newcommand{\tensionNpulsars}{24}
\newcommand{\tensionNpairs}{42}
\newcommand{\tensionNtriples}{412}
\newcommand{\tensionNflagged}{17}
\newcommand{\tensionFracFlagged}{4}
\newcommand{\tensionNpx}{42}

\begin{document}
\title{Combining Pulsar Timing data with \texttt{MetaPulsar}. \\The linearized timing model}

\date{\today}
\author{Wang-Wei Yu$^{*,\dagger}$}
\author{Rutger van Haasteren$^{\ddagger,\dagger}$}
\affiliation{Max Planck Institute for Gravitational Physics (Albert
  Einstein Institute), Leibniz Universit\"at Hannover, Callinstrasse 38,
  D-30167, Hannover, Germany}

\begin{abstract}
Combining the data of multiple pulsar timing array (PTA) collaborations increases the sensitivity to nanohertz gravitational waves (GWs) and other signals correlated between pulsars. The International PTA (IPTA) has combined such data, but historically these projects have taken many years.
We introduce \texttt{MetaPulsar}, a software package that ingests the \texttt{.par} and \texttt{.tim} files of individual data releases, and turns them into single consistent deterministic models for the combined datasets, compatible with both timing packages \texttt{PINT} and \texttt{Tempo2}.
We validate the method in four ways. First, we show for two pulsars that the linearized timing solution of a combined dataset does not depend on the parameter values in the \texttt{.par}. A mild exception is formed by the Shapiro-delay parameters. No combination considered requires a manual refit. Second, for PSR J0613$-$0200 we show that the residuals are explained by the timing model if we take intrinsic red-noise and dispersion-measure processes into account, and that the single-PTA timing posteriors agree with one another and with the combination, except for the parallax. Third, we quantify the consistency of the timing parameters measured by different PTAs with a two-sample tension statistic, for every pulsar that is observed by multiple PTAs. Finally, we carry out a gravitational-wave background search on the direct combination of the NANOGrav 9-yr, EPTA DR1 (v2.2), and PPTA DR2 data releases, the constituents of IPTA Data Release 2 (DR2). The combined dataset is assembled in under two hours, and the recovered amplitude and spectral index of the common process are consistent with the published IPTA DR2 result. With \texttt{MetaPulsar}, a joint PTA analysis can be prototyped as soon as the individual collaborations publish their data, instead of after a years-long reprocessing effort.

\end{abstract}
\maketitle
\begingroup
\renewcommand{\thefootnote}{}%
\footnotetext{\textsuperscript{*}wangwei.yu@aei.mpg.de}%
\footnotetext{\textsuperscript{$\ddagger$}rutger@vhaasteren.com}%
\footnotetext{\textsuperscript{$\dagger$}These authors contributed equally to this work.}%
\endgroup

\section{Introduction}
\label{sec:intro}

Pulsar timing arrays (PTAs) search for nanohertz gravitational waves (GWs) by monitoring the pulse times of arrival (TOAs) of millisecond pulsars across the sky \citep{HellingsDowns1983}. Now that the existence of merging black hole binaries has been confirmed by direct detection \citep{abbott2016observation}, it is expected that unresolved supermassive black hole binaries throughout the universe contribute to a stochastic gravitational-wave background (GWB) \citep{rajagopal1994ultra,jaffe2003gravitational,wyithe2003low,sesana2008stochastic,rosado2015expected}. The PTA collaborations have reported increasing evidence for such a background in their data, and they are slowly approaching the golden $5\sigma$ threshold for a detection of the GWB \citep{antoniadis2023second,agazie2023nanograv,reardon2023search,xu2023searching,miles2025meerkat}. Combining the observations of multiple observatories and PTA collaborations improves sky coverage, cadence, and radio-frequency coverage. This allows more detailed and specific noise models to be constructed, and it increases the sensitivity to the GWB and to continuous sources \citep{siemens2013stochastic}. The International PTA (IPTA) Data Release 2 (DR2) \citep{perera2019international,antoniadis2022international} demonstrated these gains, and similar improvements can be expected for upcoming IPTA datasets.

Although the IPTA has extensive experience with combining the data of the individual PTAs into a single dataset, doing so has historically required years of careful manual re-processing, taking up valuable research time and delaying scientific progress. In fact, the IPTA DR2 was delayed by so many years that it was already outdated by the time it was released: NANOGrav had published newer data~\citep{arzoumanianNANOGrav11yearData2018} before IPTA DR2 came out, and its next release showed hints of the now-accepted common red process shared among pulsars~\citep{arzoumanianNANOGrav125Yr2020} before the IPTA DR2 GWB analysis appeared, scooping the IPTA data combination effort. A main challenge in the IPTA data combination projects is their scope. Out of an abundance of thoroughness, many decisions that were already made by the individual PTAs are re-considered at the IPTA level, and the whole process is carried out twice with different software stacks. Such a project involves a lot more than plainly combining the data: it is a full data release. Joint analyses can be accelerated by using the ``Lite'' approach of selecting one dataset per pulsar \citep{larsen2025lite}, or by using the ``FrankenStat'' approach of concatenating every dataset while keeping a separate timing model for each \citep{wright2026frankenstat} (which is now also possible with \texttt{MetaPulsar}). The combination we automate below is the merged timing model. In this work we argue that the process can be restructured to be far more efficient: responsibilities can be delegated to the individual PTAs, the actual combination of the data can be automated, the skipping of certain steps can be given both theoretical and experimental grounding, and the remaining steps can be reframed as data quality checks. Gravitational-wave searches of all kinds and (advanced) noise modeling efforts can in principle be done at a later stage, independently of a data release, as often happens already in the PTA community.

As part of this work we present \texttt{MetaPulsar}, a software package and \texttt{enterprise}-compatible pulsar class that automates the direct combination of existing PTA data releases by constructing a single, timing-package-consistent deterministic model across all datasets. \texttt{MetaPulsar} solves the data combination stage of the IPTA pipeline without a manual re-assessment of the timing model and without the IPTA having to produce a new curated data release: it starts from the astrophysical models that are already present in the individual PTA releases, and it makes those models mutually consistent. We modify the timing model, especially the binary models, in very limited fashion where necessary; extensions informed by community discussions to more flexibly do this are planned. We demonstrate the data combination approach by carrying out a gravitational-wave background search on the direct combination of the individual NANOGrav 9-yr, EPTA DR1, and PPTA DR2 data releases, the constituents of IPTA DR2. The combined dataset yields a common-process posterior that is consistent with the official IPTA DR2 result, at a small fraction of the effort. \texttt{MetaPulsar} thus enables immediate ``direct'' combinations of newly released PTA datasets at negligible overhead. The package is publicly available at \url{https://github.com/vhaasteren/metapulsar}, and it can readily be used on existing datasets. Indeed, \texttt{MetaPulsar} has already been used in practice to combine the currently public data of five PTA collaborations into a single dataset for a stochastic GWB search \citep{yuStochasticGravitationalwaveBackground2025}, a combination that does not yet exist as an official IPTA data release.

This paper is part of a larger effort to address the direct combination of PTA data and spin-off analyses such as non-linear analysis of timing models of (combined) datasets. In the present paper we develop the combination method itself, and we validate it within the standard PTA Gaussian likelihood that underlies essentially all published stochastic analyses, where the timing model is linearized about the timing solution declared in the \texttt{.par} file. The organization of this paper is as follows. In Sec.~\ref{sec:theory} we review pulsar timing in the context of data combination, including the timing-parameter marginalization that is standard in PTA stochastic analyses. In Sec.~\ref{sec:metapulsar} we introduce the direct data combination approach as implemented in \texttt{MetaPulsar}: parameter alignment, astrophysical consistency, and timing package convention consistency, and we summarize the cross-package validation tests we have carried out. That section closes, in Sec.~\ref{sec:proposedworkflow}, with the workflow that we propose for future IPTA data combination efforts. In Sec.~\ref{sec:singlepulsartests} we validate the procedure pulsar by pulsar on the released data: the combined model accounts for the combined data, the individual PTA datasets of a pulsar agree with one another, and the posterior from the linearized timing model, the usual first-order expansion of the residuals about the \texttt{.par}-file solution (Sec.~\ref{sec:linearized}), does not depend on the point at which that expansion is made, except for the Shapiro-delay parameters, for the two pulsars tested. In Sec.~\ref{sec:consistency} we quantify the cross-PTA consistency of the timing parameters for every pulsar that more than one PTA observes. In Sec.~\ref{sec:iptadr2gwb} we search the direct combination of the IPTA DR2 constituents for a GWB and compare it with the published IPTA DR2 result. Throughout this paper we work with that linearized likelihood; analyses that go beyond it are deferred to later papers in this series.

\section{Background}
\label{sec:theory}

\subsection{Deterministic timing model}
\label{sec:deterministicsum}
In pulsar timing, the propagation of the pulses from the neutron star to the solar-system barycenter (SSB) is modeled as a sum of astrophysical delays with relativistic corrections, evaluated on a chosen time standard and Solar-system ephemeris \citep{hobbsTEMPO2NewPulsar2006, lorimerHandbookPulsarAstronomy2004}. With $t_{\rm arr}$ the topocentric TOA and $t_{\rm em}$ the emission time at the pulsar, the standard timing model reads
\begin{equation}
\begin{split}
t_{\rm arr} \;=\;& t_{\rm em}
\;+\; \Delta_{\rm SSB}(\hat{\mathbf{n}}, \mathrm{EPHEM})
\;+\; \Delta_{\rm bin}(\bm{\theta}_{\rm bin}) \\
&\;+\; \Delta_{\rm DM}(\mathrm{DM}, \nu)
\;+\; \cdots,
\end{split}
\label{eq:delays}
\end{equation}
with $\hat{\mathbf{n}}$ the unit vector from the SSB to the pulsar, $\mathrm{EPHEM}$ the Solar-system ephemeris, $\bm{\theta}_{\rm bin}$ the binary parameters, $\Delta_{\rm SSB}$ the Roemer, Einstein, and Shapiro terms, $\Delta_{\rm bin}$ the binary contributions when applicable~\citep{damourGeneralRelativisticCelestial1986}, and $\Delta_{\rm DM}=\mathcal{K}\,\mathrm{DM}/\nu^2$ the cold-plasma dispersion delay at radio frequency $\nu$, where $\mathcal{K}$ is the dispersion constant and $\mathrm{DM}$ the dispersion measure. Detector-specific terms come on top of these, such as per-backend offsets (\texttt{JUMP}) and frequency-dependent template coefficients (\texttt{FD*}), and those must be retained as per data-subset delays. Together with a Taylor expansion of the rotational phase about a reference epoch, Eq.~\eqref{eq:delays} defines the timing model: a fully deterministic mapping $\bm{\beta} \mapsto \Phi(t_i,\bm{\beta})$ from the timing model parameters $\bm{\beta}$ to the predicted rotational phase at the $i$-th TOA. We do not need more detail than this in the present paper. The timing model formulas are catalogued in the literature~\citep{lorimerHandbookPulsarAstronomy2004,edwardsTempo2NewPulsar2006}, the mapping is evaluated in practice with a timing package such as \texttt{Tempo2}~\citep{hobbsTEMPO2NewPulsar2006} or \texttt{PINT}~\citep{luoPINTModernSoftware2021}.

The timing residual for the $i$-th TOA relates the predicted phase $\Phi(t_i,\bm{\beta})$ to the data via
\begin{equation}
r_i \;=\; \frac{\Phi(t_i,\bm{\beta}) - N[\Phi(t_i,\bm{\beta})]}{F_0},
\qquad
F_0 \equiv \frac{{\rm d}\Phi}{{\rm d}t}\;\;,
\label{eq:phaseeq}
\end{equation}
where $N[\Phi]$ is the nearest integer rotation count, and $F_0$ is the instantaneous pulse frequency. The absolute rotational phase constitutes a gauge freedom of the timing model: only phase differences, or equivalently pulse arrival times relative to an arbitrary fiducial pulse, are observable. A constant but unknown phase offset per dataset is therefore always part of the timing model, even when it is not explicitly listed as a parameter.
Noise parameters, including scaling factors of measurement noise (e.g. \texttt{EFAC}), are not strictly part of the timing model and do not affect calculated physical delays. The residuals $r_i(\bm{\beta})$ and their partial derivatives with respect to $\bm{\beta}$ are all that the analyses of this paper require from a timing package.

\subsection{Timing packages and the \texttt{MetaPulsar} philosophy}
\label{sec:metapulsarphilosophy}

\texttt{Tempo2} and \texttt{PINT} implement the same verified physics, and on typical timing models the two packages agree to $\sim 1\,\rm{ns}$. They are, however, independent implementations with different histories and different design decisions, and some default settings and timing model conventions differ between them. The consequence that matters for data combination is that while a \texttt{.par} file obeys a common format it is not, by itself, a full unambiguous definition of the timing model. Alongside the numerical parameter values, a \texttt{.par} file carries conventions that define what those values mean: with respect to what unit of time and what time standard the parameters are measured (\texttt{UNITS}, \texttt{CLOCK}), in what coordinate system the astrometry is expressed (\texttt{POSEPOCH}, \texttt{ECL}, \texttt{T2CMETHOD}), and so on. Not all of these settings are always explicitly listed, and the two packages fill in the missing settings with different defaults. The \texttt{.tim} files have the same problem at the level of formatting and interpretation, to the point that not every published file can be read by both packages without modification. Consequently, PTA data releases are always published with the intention that they are processed with the timing package that was used to produce the dataset.

We have encountered these issues when reading historical and current published datasets by the PTA collaborations, and we believe that a dataset released by a PTA collaboration should indeed always be processed by the timing package that was used to produce it. We have developed \texttt{MetaPulsar} with that idea as a guiding principle: ingest the datasets as they are released, process each with the timing package that the data was tested on, and make only \emph{minimal} modifications to the in-memory \texttt{.par} and \texttt{.tim} files, so that \texttt{PINT} and \texttt{Tempo2} agree on the definitions of the timing model parameters and the data can be independently and coherently processed by any timing package. Where modification of input files is necessary, \texttt{MetaPulsar} will not change the physics\footnote{unless specifically requested, such as with policy regarding binary model conversions}, with cross-package agreement to within a few nanoseconds (Sec.~\ref{sec:roundtripconsistency}).

Although we do provide functionality to combine PTA datasets into single combined \texttt{.par} and \texttt{.tim} files that can be read with either package (similar to an IPTA data release), we focus most of our efforts on the approach where the datasets are kept separate and each is processed by its own timing package, with only the shared physical model made consistent between them. This makes the methodology we present future (and past!) proof: \texttt{MetaPulsar} only processes the physical model that is shared between PTAs, whereas any detector-specific modeling is allowed to stay timing-package specific. We primarily position \texttt{MetaPulsar} as a ``meta timing package'' that can do anything a timing package can do with a python interface, but where \texttt{MetaPulsar} simply uses the already-existing timing packages that the data were intended for under the hood.

\subsection{Timing model vs parameter values}
\label{sec:timingmodelvalues}

PTA projects provide the data in the form of \texttt{.tim} files. The values of the parameters in the \texttt{.par} files in principle follow from the content of the \texttt{.tim} files: when jointly analyzing PTA data, the \texttt{.tim} files carry all the information except for the definition of the model itself. The numerical values of the parameters in the \texttt{.par} files are redundant information if one has access to the \texttt{.tim} files. What the \texttt{.par} files provide that is not redundant is the model choice itself: the domain knowledge of the details of the data-taking process, and the (binary) model branch that has been chosen to model the pulsar with. Once the model and prior assumptions are fully fixed, the \emph{posterior distribution} over the model parameters mechanically follows from the data through the process of (Bayesian) inference. To do such Bayesian inference in practice with current tools, we only need the integer rotation count of the pulsar at each observed TOA and the timing model family as defined in the \texttt{.par} file. It is not necessary to precisely know the numerical value of the timing model parameters in the \texttt{.par}. This somewhat contrasts the more traditional view on data releases and pulsar timing data products, where a lot of effort is spent on how the timing residuals look and what the values of the parameters in the \texttt{.par} file are. While we do not claim that there is no scientific use for releasing proper \texttt{.par} files with values of parameters to the best of our ability, we emphasize in this paper on \texttt{MetaPulsar} that the values of the timing model parameters in the \texttt{.par} file do not affect the results when the timing model is evaluated in full. We find in Sec.~\ref{sec:singlepulsartests} that for the linearized likelihood used in this paper this is also true for the astrometric, spin, and Keplerian parameters and for the red-noise and dispersion-measure posteriors of the two pulsars we test, but not entirely for the Shapiro-delay parameters (Fig.~\ref{fig:j0613_triple}). For a timing model evaluated in full, the statement covers single-pulsar timing analysis, interstellar medium studies, and GW analysis alike. Only if one uses the residuals or \texttt{.par} file parameter values directly in isolation would there be a difference, but that ignores the large covariances that are present in the noise and timing models, so such analyses would be unreliable regardless of the \texttt{.par} file optimization procedure.

Getting that posterior, for a fully specified timing model and a given set of observations, is a standard analysis in the literature, usually called single pulsar noise and timing analysis (SPNTA). Over a decade ago there were the \texttt{TempoNest}~\citep{lentatiTemponestBayesianApproach2014} and NANOGrav~\citet{vigelandBayesianInferencePulsartiming2014} studies, and more recently the \texttt{enterprise}-based work of \citet{kaiserGeneralizedNonlinearBayesian2026a} and the more modern \texttt{Vela.jl} work~\citep{susobhananBayesianPulsarTiming2025b}, which describe how to do this in a Bayesian framework end-to-end. When the timing model is evaluated in full, as these methods do, they obtain the \emph{exact} same posterior distribution of the timing model parameters, regardless of the values in the \texttt{.par} file that was ingested.

\subsection{Linearized PTA likelihood and the design matrix}
\label{sec:linearized}
Standard PTA stochastic analyses operate on a linearized version of the deterministic timing model. Let $\bm{\beta}_0$ denote the nominal values of the timing model parameters $\bm{\beta}$ in a given \texttt{.par} file. Linearizing the residuals about $\bm{\beta}_0$ yields
\begin{equation}
\mathbf{r}(\bm{\beta}) \;\approx\; \mathbf{n} \;+\; \mathbf{M}\,\bm{\varepsilon},
\qquad
\bm{\varepsilon} \equiv \bm{\beta}-\bm{\beta}_0,
\label{eq:linear}
\end{equation}
where $\mathbf{M} \equiv \partial \mathbf{r}/\partial \bm{\beta}\big|_{\bm{\beta}_0}$ is the design matrix, and we use $\mathbf{n}$ to denote the sum of all stochastic contributions, such as (white) measurement noise, intrinsic red noise, DM variability, and any GW signal or other stochastic process. The entries of $\mathbf{M}$ are the partial derivatives of the deterministic model, evaluated at the nominal timing solution. This linearization is the standard approximation used in contemporary flagship PTA results. The approximation is highly accurate when the timing solution is already close to the relevant likelihood maximum, so that the remaining timing offsets are small \citep{van2013understanding}.

The use of a linearized PTA likelihood is somewhat in tension with the view of Sec.~\ref{sec:timingmodelvalues}, because the linearization represents the tangent space of timing solution at a specific point in parameter space: where $\bm{\beta}=\bm{\beta}_0$. This means that the linearization is only valid in a small neighborhood of the nominal timing solution, and the linearization is not valid for the entire parameter space. In \texttt{MetaPulsar} a lot of care is taken to make sure the data combination and the resulting PTA analysis fully capture the representative physics down to the level of a few nanoseconds. Where the linearization partially fails for the combined datasets is shown in Sec.~\ref{sec:singlepulsartests}.

\subsection{Noise model}
\label{sec:noisemodel}

The sum of all stochastic processes $\mathbf{n}$ in Eq.~\eqref{eq:linear} is typically modeled as a sum of zero-mean Gaussian processes (GPs). While the noise model is in principle not a necessary component of the timing model, and separate from the process of data combination, we will run tests on public datasets that make use of the noise model in this work. We therefore briefly describe the most common GP components that are used in PTA analyses.

The most common GP components are the white noise and the epoch-correlated noise. For the $i$-th TOA, the formal template-matching uncertainty $\sigma_i$ and a \texttt{Tempo2} \texttt{EQUAD} $Q_i$ are added in quadrature and then rescaled by a multiplicative \texttt{EFAC} $E_i$,
\begin{equation}
    \sigma_{i,\mathrm{eff}}^2
    \;=\;
    E_i^2\,\bigl(\sigma_i^2 + Q_i^2\bigr) ,
    \label{eq:whitenoise}
\end{equation}
where both parameters are piecewise-constant across backend groups. A second time-uncorrelated contribution, \texttt{ECORR}, captures epoch-to-epoch jitter that is correlated across the frequency sub-bands within the same observing epoch. Its covariance is $C_q = \sum_{e} q_e^2\,\mathbf{v}_e\,\mathbf{v}_e^{\!\top}$, where $e$ indexes the epochs, $\mathbf{v}_e$ selects the TOAs of epoch $e$, and $q_e$ is the \texttt{ECORR} amplitude of the backend that observed epoch $e$.

Beyond these measurement terms there are time-correlated processes: intrinsic spin noise, DM variations, and any GW-induced signal. These are stationary GPs on a Fourier basis $f_k = k/T$, with bin width $\Delta f = 1/T$. The power-law piece of the spectrum is
\begin{equation}
S(f) = \frac{A^2}{12 \pi^2}\,\mathrm{yr}^{3} \left(\frac{f}{f_{\rm yr}}\right)^{-\gamma},
\label{eq:powerlaw}
\end{equation}
where $A$ is the dimensionless amplitude, $\gamma$ is the spectral index, and the reference frequency is fixed at $f_{\rm yr} = 1\,\mathrm{yr}^{-1}$. The quantity that enters the likelihood is the prior variance of each sine and cosine coefficient. For the intrinsic red noise and for the DM variations that variance is a power law plus a frequency-independent floor,
\begin{equation}
P(f)
\;=\;
S(f)\,\Delta f
\;+\;
\kappa^{2},
\label{eq:flattail}
\end{equation}
with a separate $(A, \gamma, \kappa)$ for each process. The common GW process uses the power-law term alone, $P(f) = S(f)\,\Delta f$. The consistency comparison of Sec.~\ref{sec:consistency} replaces Eq.~\eqref{eq:flattail} by a free spectrum, one amplitude $\rho_k$ in the $k$-th bin, with $P(f_k) = \rho_k^2$. Together with the white noise and \texttt{ECORR} terms, these components define the noise covariance $\mathbf{C}(\bm{\eta})$. The hyperparameters $\bm{\eta}$ and the uniform priors placed on them are listed in Table~\ref{tab:noisepriors}.

\begin{table*}
\caption{Noise hyperparameters and the uniform priors used wherever they are
sampled. Red noise and DM variations each carry their own copy of
$(A,\gamma,\kappa)$, or of $\rho_k$ when the free spectrum replaces
Eq.~\eqref{eq:flattail}. The common process has no floor $\kappa$. White noise
is sampled under these priors in the MCMC of
Fig.~\ref{fig:j0613_pta_corner} and in both comparisons of
Sec.~\ref{sec:consistency}. It is held fixed at the empirical-Bayes maximum defined in
Sec.~\ref{sec:singlepulsartests} in the other tests of that section, and in the gravitational-wave search of
Sec.~\ref{sec:iptadr2gwb}. A backend with fewer than four TOAs is not sampled:
its \texttt{EFAC} is fixed at $1$ and its \texttt{EQUAD} at $1\,\mathrm{ns}$.
\texttt{ECORR} is present only on backends that record several TOAs in one
observing epoch. Parameters with units in the $\log$ are divided by $\mathrm{s}$ (seconds).}
\label{tab:noisepriors}
\begin{tabular}{p{0.20\textwidth}p{0.12\textwidth}p{0.60\textwidth}}
\hline
Parameter & Prior & Meaning \\
\hline
$E$ & $[0.1,\,10]$ & \texttt{EFAC} of one backend, Eq.~\eqref{eq:whitenoise}. \\
$\log_{10}(Q/\mathrm{s})$ & $[-9,\,-3]$ & \texttt{Tempo2} \texttt{EQUAD}, $Q$, of one backend, in seconds. \\
$\log_{10}(q/\mathrm{s})$ & $[-9,\,-3]$ & \texttt{ECORR} amplitude, $q$, of one backend, in seconds. \\
$\log_{10} A$ & $[-20,\,-10]$ & Power-law amplitude, $A$, of the red noise or of the DM variations, Eq.~\eqref{eq:powerlaw}. \\
$\gamma$ & $[1,\,7]$ & Spectral index of that same process. \\
$\log_{10}(\kappa/\mathrm{s})$ & $[-10,\,-4]$ & Flat-tail amplitude of that process, $\kappa$ in Eq.~\eqref{eq:flattail}, in seconds. \\
$\log_{10}(\rho_k/\mathrm{s})$ & $[-10,\,-4]$ & Free-spectrum amplitude, $\rho$, in Fourier bin $k$, red noise or DM. The coefficient variance is $\rho_k^2$. \\
$\log_{10} A_{\rm GWB}$ & $[-20,\,-10]$ & Amplitude of the common process, $A_{\rm GWB}$, the power-law term of Eq.~\eqref{eq:powerlaw} only. \\
$\gamma_{\rm GWB}$ & $[1,\,7]$ & Spectral index of the common process. \\
\hline
\end{tabular}
\end{table*}

\subsection{Timing-parameter marginalization}
\label{sec:marginalization}

Standard PTA stochastic analyses do not sample the timing model parameters.
Because the likelihood that follows from Eq.~\eqref{eq:linear} is Gaussian
in the timing offsets $\bm{\varepsilon}$, those offsets can be integrated out
analytically. In principle nothing stops us from sampling $\bm{\varepsilon}$
jointly with the noise hyperparameters $\bm{\eta}$ (the linearized parameters
are ordinary model parameters), but a combined dataset can easily carry
hundreds of timing parameters per pulsar (per-backend \texttt{JUMP}s,
\texttt{FD} coefficients, \ldots). Extra parameters make the numerical sampling
less efficient, and because the analytic route is exact for linear parameters we
can always analytically marginalize them out. We place a
zero-mean Gaussian prior on the offsets, $\bm{\varepsilon} \sim
\mathcal{N}(\mathbf{0},\,\bm{\Lambda})$, centered on the nominal timing
solution. With all stochastic terms combined into
$\mathbf{n} \sim \mathcal{N}(\mathbf{0},\,\mathbf{C}(\bm{\eta}))$, the
marginalization yields
\begin{equation}
\mathcal{L}_{\rm marg}(\mathbf{r}\,|\,\bm{\eta})
\;=\;
\det\left(2\pi\mathbf{C}_{\rm t}\right)^{-1/2}
\exp\!\left[
-\tfrac12\,
\mathbf{r}^{\!\top}
\mathbf{C}_{\rm t}^{-1}
\mathbf{r}
\right],
\label{eq:gaussmarg}
\end{equation}
with
\begin{equation}
\mathbf{C}_{\rm t} \;\equiv\; \mathbf{C} + \mathbf{M}\bm{\Lambda}\mathbf{M}^{\!\top}.
\end{equation}
The interpretation of Eq.~\eqref{eq:gaussmarg} is that the linearized part of
the timing model has become one more zero-mean Gaussian process in line with 
Sec.~\ref{sec:noisemodel}, with basis $\mathbf{M}$ and prior covariance
$\bm{\Lambda}$. This is also how the marginalization is implemented in modern
PTA analysis codes, where the Woodbury identity and matrix determinant lemma,
\begin{align}
\bm{\Sigma}
&\;\equiv\;
\left(\mathbf{M}^{\!\top}\mathbf{C}^{-1}\mathbf{M} + \bm{\Lambda}^{-1}\right)^{-1},\\[0.5ex]
\mathbf{C}_{\rm t}^{-1}
&\;=\;
\mathbf{C}^{-1}
- \mathbf{C}^{-1}\mathbf{M}\,
\bm{\Sigma}\,
\mathbf{M}^{\!\top}\mathbf{C}^{-1},
\label{eq:woodbury}
\\[0.5ex]
\det \mathbf{C}_{\rm t}^{-1}
&\;\equiv\;
\det \left(\bm{\Lambda}^{-1} + \mathbf{M}^{\!\top}\mathbf{C}^{-1}\mathbf{M}\right) \det(\bm{\Lambda})\det(\mathbf{C}),
\end{align}
keeps the evaluation efficient~\citep{vanhaasterenNewAdvancesGaussianprocess2014}.
In the limit of an uninformative prior, $\bm{\Lambda}^{-1}\rightarrow\mathbf{0}$,
Eqs.~\eqref{eq:gaussmarg} and \eqref{eq:woodbury} reduce to
\begin{align}
\mathcal{L}_{\rm marg}(\mathbf{r}\,|\,\bm{\eta})
&\sim \frac{1}{\sqrt{\det\bm{\Lambda}}}
\frac{
\exp\!\left[-\tfrac12\,\mathbf{r}^{\!\top}\mathbf{P}\,\mathbf{r}\right]
}{
\sqrt{\det\left(\mathbf{M}^{\!\top}\mathbf{C}^{-1}\mathbf{M}\right)\,\det(\mathbf{2 \pi C})}
}\\[0.5ex]
&\propto
\frac{
\exp\!\left[-\tfrac12\,\mathbf{r}^{\!\top}\mathbf{P}\,\mathbf{r}\right]
}{
\sqrt{\det\left(\mathbf{M}^{\!\top}\mathbf{C}^{-1}\mathbf{M}\right)\,\det\mathbf{C}}
},
\\[0.5ex]
\mathbf{P}
&\equiv
\mathbf{C}^{-1}
- \mathbf{C}^{-1}\mathbf{M}
\left(\mathbf{M}^{\!\top}\mathbf{C}^{-1}\mathbf{M}\right)^{-1}
\mathbf{M}^{\!\top}\mathbf{C}^{-1},
\label{eq:marginal}
\end{align}
which is the familiar marginalized PTA likelihood~\citep{van2013understanding};
the proportionality absorbs the divergent but parameter-independent factor
$\det\bm{\Lambda}$.

We note that these Gaussian priors are useful in practice. Prior information on timing model parameters does exist (independent VLBI
astrometry~\citep{ding2023mspsrpi}, or posteriors from earlier datasets, for example), and
Eq.~\eqref{eq:gaussmarg} is the correct way to feed it into the analysis.
While current PTA analysis codes do not do this, we suggest that this practice
should be adopted: the code infrastructure is there, and using such priors
is a simple matter to implement. Priors also matter when the data alone constrain some direction of
parameter space poorly: in a combined dataset, detector-specific parameters of
different datasets can overlap into (nearly) degenerate combinations, for which
$\mathbf{M}^{\!\top}\mathbf{C}^{-1}\mathbf{M}$ in Eq.~\eqref{eq:marginal} is
(nearly) singular and the flat-prior expression is ill-behaved. A proper prior
keeps the marginalization well-defined in those cases. This would
address many numerical issues in current PTA analysis codes, including the move
towards full-PTA single-precision computations on GPU-based architectures.

Equation~\eqref{eq:marginal} is the likelihood on which essentially all
published PTA stochastic results are based, and it is the likelihood we use
throughout this paper. Its accuracy is governed by the same condition as
Eq.~\eqref{eq:linear}: the posterior mass has to be confined to the
neighborhood of $\bm{\beta}_0$ in which the linearization holds, which is the
case for most parameters of well-timed
millisecond pulsars~\citep[e.g.][]{agazieNANOGrav15Yr2023}. When this is not the case, the modern approach is to partition $\bm{\beta}$, keeping a subset of parameters,
typically binary parameters with manifestly non-Gaussian
posteriors, in full nonlinearity while the remainder is marginalized analytically as above.
Such an analysis is referred to as single pulsar noise and timing analysis (SPNTA), like we describe in 
Sec.~\ref{sec:timingmodelvalues}: for those parameters, the inference is
completely independent of the values printed in the released \texttt{.par} files. To support both SPNTA and the traditional fully linearized approaches we developed \texttt{nltiming}~\citep{nltiming}, an extension to the two
most-widely used PTA analysis packages, \texttt{enterprise}~\citep{ellisENTERPRISEEnhancedNumerical2020}
and \texttt{discovery}~\citep{discovery}, which is fully compatible with
\texttt{MetaPulsar} and can use \texttt{Vela.jl}~\citep{susobhananBayesianPulsarTiming2025b}
and \texttt{Tempo2} as a timing backend. In the present paper we stay in the
fully linearized limit of Eq.~\eqref{eq:marginal}; the nonlinear treatment and
its consequences for data combination are the subject of a later paper we are
preparing. Fig.~\ref{fig:j0613_triple} shows in what way results depend on
the point at which the linearization approximation is made: it depends on the
Shapiro-delay parameters, and the differences for PTA pulsars are small.

\section{Direct combination with \texttt{MetaPulsar}}
\label{sec:metapulsar}

\subsection{Fully specified timing models}
\label{sec:fullspecification}

In Sec.~\ref{sec:metapulsarphilosophy} we argued that a \texttt{.par} file is unambiguously
specified only once every convention that affects the delay calculation is
written explicitly. When a combination has to be evaluated by both
\texttt{PINT} and \texttt{Tempo2}, \texttt{MetaPulsar} produces such files by
rewriting the shared deterministic model onto a \texttt{.par} file in which
all conventions are fixed to an explicit ``common profile'' that both packages are
compatible with. This common profile is listed in
Table~\ref{tab:commonprofile}. What \texttt{MetaPulsar} actually does to the
\texttt{.par} files is a fixed recipe.
Though an optimization/fit of the timing model parameters with respect to the data is trivial, we stress up front that none of these steps requires it. Each one is a
relabeling, a unit conversion, a coordinate rotation, or a copy of values
from one file to another. The recipe, which we fully explain later this section, is:
\begin{enumerate}
  \item Before changing any \texttt{.par} file, determine the integer
    rotation count of every TOA from its released \texttt{.par}/\texttt{.tim}
    pair and write it to the \texttt{.tim} file as a \texttt{-pn} flag
    (Eq.~\eqref{eq:phaseeq}). This preserves the phase-connected solution
    established by the individual PTA; later edits, in particular removing
    \texttt{DMX}, can otherwise break phase connection.
  \item Pick the reference PTA (Table~\ref{tab:combinationinputs}). Its
    parameter values, its clock and ephemeris, and its binary model will be
    used in the combination. The other datasets are the targets.
  \item Write a hybrid orbit as an orbital frequency.
    A few files give the period as \texttt{PB} (days) and its
    derivatives as \texttt{FB1}, \texttt{FB2}, \ldots (so derivatives of the frequency \texttt{FB0} instead of the period \texttt{PB}; a combination that PINT does not support).
    The \texttt{PB} line is replaced by
    \texttt{FB0} $= 1/(\texttt{PB}\times 86400\,\mathrm{s})$,
    with the fit flag copied and the uncertainty transformed;
    the \texttt{FBn} lines are unchanged.
    Both packages then share one free parameter, the orbital frequency.
    Files already written entirely as \texttt{PB} or as \texttt{FB0} are left as they are.
  \item Give every binary model an explicit name. A \texttt{BINARY T2}
    line is replaced by the name of the model that its parameters select,
    such as ELL1, ELL1H, DD, or DDK (App.~\ref{app:binarymodels}). For DDK
    the Kopeikin angles are mapped from the IAU to the DT92 convention, and
    a \texttt{SINI} line is dropped because DDK already derives $\sin i$
    from the inclination (Table.~\ref{tab:binaryparams}).
  \item Expand the aggregate \texttt{TEMPO1} switch into the six settings it
    stands for.
  \item Remove the deterministic terms that one of the two packages cannot
    evaluate, such as \texttt{WAVE}, \texttt{IFUNC}, glitch and
    chromatic-index terms, \texttt{DMMODEL}, telescope-specific corrections,
    second derivatives of the proper motion, and the non-constant solar-wind
    terms. The complete list is maintained in the software, and each removal
    is reported together with the PTA and the parameter names. The policy
    can also be set to refuse the combination instead.
  \item Bring every file to explicit \texttt{UNITS TDB}. A file without a
    \texttt{UNITS} line is interpreted the way its own package interprets
    it, i.e.\ TDB for \texttt{PINT} and TCB for \texttt{Tempo2}, and a TCB
    file is converted to TDB by its own timing package. We note that at the time
    of this writing \texttt{Tempo2} does not define \texttt{TDB} according to
    the IAU 2006 or 2010 resolutions, but uses its own definition of
    \texttt{TDB} with a very slightly different clock rate. The difference is
    small, but we argue this should be addressed in a future version of
    \texttt{Tempo2}.
  \item Rotate ecliptic astrometry numerically, through ICRS, onto the
    \texttt{IERS2003} obliquity. Equatorial models lose their \texttt{ECL}
    line.
  \item Write the common profile of Table~\ref{tab:commonprofile} into
    every file.
  \item Copy the astrometric, spin, binary, and dispersion parameters of
    the reference onto every target, values included. In other words: a
    target also inherits the binary model of the reference, and any
    parameter of these components that its own release did not carry,
    whereas a parameter that only the target carried is dropped. For an
    ELL1H model with $h_3$ and $h_4$ this is also the point where both
    packages are given the same number of harmonics
    (Table~\ref{tab:commonprofile}).
  \item Clean up the dispersion: remove \texttt{DMX}, keep each PTA's own
    \texttt{DM} and make sure it is free, fix \texttt{DMEPOCH} to the
    reference value, and set \texttt{DM1} and \texttt{DM2} to zero as free
    parameters, adding them if the file does not have them, if the
    conventional ``diagonal Phi'' models are
    used~\citep{crisostomiDiagonalApproximationsImproved2025}.
  \item Where \texttt{PINT} and \texttt{Tempo2} would still evaluate the
    shared binary model differently, convert it. This is only an issue for
    the ELL1 family, because the two packages truncate the expansion in $e$
    at a different order. We therefore estimate the size of the neglected
    terms, and where that estimate exceeds $1\,\mathrm{ns}$ the model is
    converted to DD or DDH, after which the predicted delays before and
    after the conversion are compared (Sec.~\ref{sec:binarymodels},
    App.~\ref{app:binarymodels}).
\end{enumerate}
The detector-specific parameters, i.e.\ the \texttt{JUMP}s, the
\texttt{FD} coefficients, and the \texttt{EFAC}/\texttt{EQUAD}/\texttt{ECORR}
lines, keep their values. The only thing that happens to them is a change of
spelling where the two packages would otherwise read the same line
differently: a white-noise line that selects its TOAs by flag is written as
\texttt{T2EFAC}, \texttt{T2EQUAD}, or \texttt{TNECORR}, which both packages
understand, because \texttt{Tempo2} interprets a bare \texttt{EFAC -f
\ldots} as a global scalar. No arrival time, TOA uncertainty, or existing
flag is modified. Step~1 is performed on the released files before any
\texttt{.par}-file edit. The remaining order matters: steps~3--9 have to run
before step~10, otherwise the copy would carry a convention from the reference
that the target reads differently.

That order also makes clear what the role of the reference PTA really is.
Steps~3--9 make sure that every parameter has the same meaning in both
packages and in every file, and only then does step~10 decide which model is
written into all of them. Taking that model from the reference PTA is the
natural default, but it is not a requirement: once the files agree on their
conventions, the model that is copied in step~10 can just as well be one that
the user supplies, and swapping out the binary model in particular is a
simple operation. In the current version of \texttt{MetaPulsar} the model is
simply inherited from the reference PTA; in a future version the user will be able to supply different models.
At the end of this we have: a) a \texttt{.par} file
where both \texttt{Tempo2} and \texttt{PINT} agree to within a few nanoseconds
on \emph{the astrophysical} model (Sec.~\ref{sec:roundtripconsistency}), and
b) \texttt{.par} files of a pulsar from various PTAs where all timing model
parameters have the same meaning. Once (a) and (b) are satisfied, we can
``lock'' the timing model parameters between PTAs, and treat the joined
astrophysical model as a single timing model.

\begin{table}
\caption{The common convention profile that \texttt{MetaPulsar} writes into
the \texttt{.par} files of a combination that is evaluated by both timing
packages.}
\label{tab:commonprofile}
\begin{tabular}{l>{\raggedright\arraybackslash}p{0.56\columnwidth}}
\hline
Setting & Written value \\
\hline
\texttt{UNITS} & \texttt{TDB} \\
\texttt{CLK} & clock realization of the reference PTA \\
\texttt{EPHEM} & ephemeris of the reference PTA \\
\texttt{T2CMETHOD} & \texttt{IAU2000B} \\
\texttt{TIMEEPH} & \texttt{FB90} \\
\texttt{DILATEFREQ} & \texttt{N} \\
\texttt{CORRECT\_TROPOSPHERE} & \texttt{N} \\
\texttt{PLANET\_SHAPIRO} & \texttt{N} \\
Solar-system Shapiro delay & on (\texttt{NO\_SS\_SHAPIRO} removed) \\
\texttt{SWM} & \texttt{0} \\
\texttt{IPM} & \texttt{1}, only for datasets processed with \texttt{Tempo2} \\
\texttt{NE\_SW} & value of the reference PTA if its file states one, otherwise $4\,\mathrm{cm}^{-3}$ \\
\texttt{ECL} & \texttt{IERS2003}, for models in ecliptic coordinates \\
\texttt{NHARM}, \texttt{NHARMS} & for ELL1H with $h_3$ and $h_4$: the largest harmonic count that any input declares, and at least 7, written under both names; removed for $h_3$ with $\varsigma$ \\
\hline
\end{tabular}
\end{table}

Note that by default \texttt{MetaPulsar} only runs the convention-alignment
operations in steps~5--9 and step~12 when the combination actually contains
both timing packages. A pulsar that
is observed by a single PTA, or by several PTAs whose data were all
processed with the same package, keeps its native units, its \texttt{ECL},
and its switches, because there is no second package to align against. In
this paper we force the full profile on every combination, including the
pulsars that are observed by a single PTA, because every combination is
evaluated by both timing packages (Table~\ref{tab:combinationinputs}). We
stress that the values in Table~\ref{tab:commonprofile} are not a
recommendation for analyses that use only one timing package. They are the
settings for which we have verified that the two packages agree.

The clock realization and the ephemeris are copied from the reference PTA.
They are not fixed constants, but follow from the choice of reference PTA in
Table~\ref{tab:combinationinputs}. The clock is written with the keyword
\texttt{CLK}: \texttt{Tempo2} reads \texttt{CLK} and silently ignores
\texttt{CLOCK}, in which case it falls back to its default realization,
whereas \texttt{PINT} accepts \texttt{CLK}. A \texttt{TT(BIPM)} realization
without a year is ambiguous, and it is not accepted.

The solar-wind density \texttt{NE\_SW} is written explicitly, because the two
packages disagree when that line is absent: \texttt{Tempo2} then applies
$4\,\mathrm{cm}^{-3}$, and \texttt{PINT} applies none. We take the value from
the file of the reference PTA if that file states one (as \texttt{NE\_SW},
\texttt{SOLARN0}, or \texttt{NE1AU}), and we use $4\,\mathrm{cm}^{-3}$
otherwise, provided that a \texttt{Tempo2} dataset is part of the
combination. A \texttt{PINT}-only combination without such a line is left
without one, since the \texttt{PINT} default is zero and there is nothing to
align. A different value in the file of another PTA is overwritten.
Solar-wind terms that are not constant are removed, since we consider the
stochastic solar wind to be part of the noise model. \texttt{IPM} is a
\texttt{Tempo2} switch, so it is only written for datasets that are processed
with \texttt{Tempo2}.

The ecliptic rotation of step~8 recomputes the coordinates as well as the
\texttt{ECL} label. Changing the label alone would move the pulsar on the
sky. An equatorial model has no obliquity, so its \texttt{ECL} line is
removed if equatorial coordinates are used.

Of the settings stored in \texttt{TEMPO1}, the profile changes one:
\texttt{T2CMETHOD TEMPO} is written as \texttt{T2CMETHOD IAU2000B}.
The binary model is not part of this profile
(Sec.~\ref{sec:binarymodels}).

The combination is mechanical and fully deterministic. Because we need to read with both \texttt{PINT} and \texttt{Tempo2} as part of this manuscript, we have built an option into \texttt{MetaPulsar} to produce ``canonical'' \texttt{.tim} files that can be read by both timing packages. This forces the common profile, even on pulsars that are observed by a single PTA. Table~\ref{tab:combinationinputs} lists all inputs and decisions that \texttt{MetaPulsar} accepts, together with the defaults, which are the values
used for every combination in this paper except for the two alignment-policy
settings named in the caption. There are no other inputs: given the same \texttt{.par} and \texttt{.tim} files and the same decisions, the processing always produces the same combined pulsar. No arrival time,
TOA uncertainty, or existing flag is modified in the process. The canonical
\texttt{.tim} files that we write out for inspection only change how the
same TOAs are written down: the \texttt{INCLUDE}s are flattened into a
single file, a \texttt{TIME} offset is folded into the printed MJD, and
every TOA is flagged with its PTA, which in turn allows a \texttt{JUMP} on an
MJD window to be written as a \texttt{JUMP} on that flag.

\begin{table*}
\caption{The complete set of inputs and decisions of a \texttt{MetaPulsar}
combination. The defaults are the values used throughout this paper, except
that every combination is evaluated by both timing packages, and a source
that carries $h_3$ alone is given a prior on $\varsigma$ rather than
refused. They encode the recommendations of the main text. The processing that each row
controls is the recipe of Sec.~\ref{sec:fullspecification}.}
\label{tab:combinationinputs}
\begin{tabular}{p{0.20\textwidth}p{0.74\textwidth}}
\hline
Input / decision & Description \\
\hline
\texttt{.par}, \texttt{.tim} files & The released timing models and TOAs of
each PTA, ingested as published. Required; located by the release-layout
discovery or supplied explicitly. \\
combination strategy & How the astrophysical model is treated across PTAs.
Default \texttt{shared}, which merges it; \texttt{per\_pta} composes the
releases unchanged. \\
reference PTA & The dataset whose parameter values and conventions are copied
onto the others. Default: the first one listed. \\
shared components & The timing model components that are merged across PTAs.
Default: astrometry, spindown, binary, and dispersion. \\
excluded parameters & Parameters kept PTA-local even when their component is
shared. Default: \texttt{DM} (see the text). \\
DM derivatives & Whether free \texttt{DM1} and \texttt{DM2} parameters are
added to the model, at value zero. Default: yes. \\
pulse numbers & Whether every TOA carries its pinned integer rotation count
(\texttt{-pn}; cf.\ Eq.~\eqref{eq:phaseeq}), so that all timing packages
agree on the pulse assignment. Default: yes. \\
alignment policy & The decisions in steps 6, 9, and 12 of the recipe of
Sec.~\ref{sec:fullspecification}, which are made so that \texttt{Tempo2} and
\texttt{PINT} agree on the timing model: whether the common profile is
written only for combinations that contain both packages or always, whether
unsupported terms are stripped or refused, whether an ELL1-family model is
converted at a disagreement of $>1\,\mathrm{ns}$, and what is done with a
source that carries $h_3$ alone. Default: the profile for mixed-package
combinations only, strip with warnings, convert at $>1\,\mathrm{ns}$, and
refuse $h_3$ alone. \\
clock directory & The clock-correction realization handed to the timing
packages. Default: taken from the environment. \\
output directories & Where the processed \texttt{.par} and canonical
\texttt{.tim} files are written for inspection. Default: not written. \\
\hline
\end{tabular}
\end{table*}

The most impactful choice in Table~\ref{tab:combinationinputs} is the reference PTA: its parameter
values, its conventions, and its binary model become those of the combined
dataset (Sec.~\ref{sec:binarymodels}). In an ideal world, all PTAs would simply use a full general relativistic binary model in combination with an astrometry and spin model. Unfortunately, modern PTAs do not always do this consistently, and we need to allow the PTAs to make scientific modeling choices. The reference PTA is therefore user input.
In the current version of
\texttt{MetaPulsar} the combined timing model is inherited from the reference PTA. Future versions will give the user more control over the
precise timing model, especially the binary model, by replacing the
reference PTA's model with a user-provided one before the processing runs.
The alignment policy contains the remaining expert decisions. Its defaults resolve everything from the reference PTA, so in practice the user only provides a list of \texttt{.par}/\texttt{.tim} files and the name of the reference PTA. The processed per-PTA \texttt{.par} files and
the canonical \texttt{.tim} files can be written to disk, so every
combination can be reproduced and inspected outside of \texttt{MetaPulsar}.
The released datasets do not describe dispersion in the same way. NANOGrav
uses \texttt{DMX}, whereas the releases that were made with \texttt{Tempo2}
use \texttt{DMMODEL} or a Taylor series in the DM. In a shared combination
these descriptions cannot be used side by side. We therefore remove the
\texttt{DMX} and \texttt{DMMODEL} parameters. Each PTA keeps its own
\texttt{DM} as a free parameter, at the value in its released file.
\texttt{DMEPOCH} is fixed to the value of the reference PTA, and free
\texttt{DM1} and \texttt{DM2} parameters are added at value zero and shared
between the PTAs, so that all datasets are linearized with respect to the same
secular trend in the dispersion. The stochastic DM variations are modeled with
the GP of Sec.~\ref{sec:noisemodel}, and they are not part of the
deterministic model.

The reason that \texttt{DM} is kept local to each PTA is the same effect that
the PTA collaborations nowadays model with \texttt{FDJUMPDM}. The TOAs were
formed from the pulse profiles with a specific value of the DM. Using a
different value of the DM in the timing model does not fully undo that, and some
frequency-dependent mismatch remains. \texttt{FDJUMPDM} was introduced by the IPTA as a calibration parameter to partially account for that effect: it is an offset in DM per
dataset, so that each subset keeps the value it was formed with, while the
pulsar still has a single astrophysical DM. Keeping the \texttt{DM} of each
PTA as its own free parameter has the same effect. In the per-PTA
\texttt{.par} files this is the local \texttt{DM}. In the single combined
\texttt{.par} file that we use for the \texttt{PINT}-only and
\texttt{Tempo2}-only GWB analyses of Sec.~\ref{sec:iptadr2gwb}, it is written
as an \texttt{FDJUMPDM}, equal to the \texttt{DM} of that PTA minus the
\texttt{DM} of the reference PTA.

\subsection{Cross-package consistency}
\label{sec:roundtripconsistency}

Whether the rewritten timing models are indeed fully specified can be checked by comparing the calculated residuals with both \texttt{PINT} and \texttt{Tempo2}.
We first verified this on simulated data for all PTA datasets publicly released since
2015, and then confirmed it on the three data releases that are combined in this
paper: EPTA DR1, the NANOGrav 9-yr release, and PPTA DR2, together $95$
pulsar datasets and $2.0\times10^{5}$ TOAs. For each dataset,
\texttt{MetaPulsar} writes the processed timing model back out as a standalone
pair of \texttt{.par} and \texttt{.tim} files, both packages load that pair
without a fit, and the two residual series are compared on the actual released
TOAs. The median RMS difference over the $95$ datasets is $1.2\,\mathrm{ns}$,
and all $95$ lie below $2\,\mathrm{ns}$. The largest disagreement is
$1.8\,\mathrm{ns}$ (the NANOGrav 9-yr dataset of PSR~J2017$+$0603). The
combined \texttt{.par}/\texttt{.tim} pairs of the $62$ pulsars pass the same
test with a median of $1.3\,\mathrm{ns}$ and a maximum of $1.8\,\mathrm{ns}$
(App.~\ref{app:roundtrip}). These differences are well below the TOA uncertainties.

As a part of our efforts to compare \texttt{PINT} and \texttt{Tempo2}, we stumbled upon many implementation and software issues that needed to be corrected. For each of these, we have filed the respective pull requests. See App.~\ref{app:softwarefixes}.

\subsection{Binary models}
\label{sec:binarymodels}
The released \texttt{.par} files of one pulsar often  do not use the same binary
model. The PTAs choose among several parametrizations of the same orbit, each
adequate at the sensitivity of its own dataset, and the EPTA and PPTA releases
declare most of their binaries with the generic T2 model of \texttt{Tempo2}.
The T2 model is a decision tree rather than a single model: depending on which parameters appear in the file, it selects among several delay formulae. \texttt{PINT} does not implement it. In a combined dataset we need a single binary model. The binary orbit is astrophysics, so it belongs to the shared
part of the timing model (Sec.~\ref{sec:fullspecification}), and exactly one
explicitly named binary model must describe it so that it is uniquely defined, across all PTAs and in both
timing packages.

\texttt{MetaPulsar} does this in three steps (App.~\ref{app:binarymodels}). First, it uses \texttt{PINT} to resolve every T2 declaration to the concrete model that the parameters select. Second, it gives all datasets of the pulsar the binary model of the reference PTA. Third, where \texttt{PINT} and \texttt{Tempo2} evaluate that model
differently, which happens within the ELL1 family at the nanosecond level, it
converts the shared model to DD or DDH, which both packages evaluate
identically. We stress that none of these steps is a refit. Resolving and converting leave the
predicted delays unchanged, the conversion is verified per pulsar by comparing
the delays before and after, and the round trip of
Sec.~\ref{sec:roundtripconsistency} covers the result as a whole. The binary
parameters reported in Secs.~\ref{sec:singlepulsartests}
and~\ref{sec:consistency} are therefore those of the combination, which need
not be the ones printed in a released \texttt{.par} file
(Table~\ref{tab:binaryparams}).

\subsection{Proposed IPTA workflow}
\label{sec:proposedworkflow}

The IPTA data combination projects have historically bundled many distinct
tasks into a single, monolithic effort. The TOAs, the timing models, the parameter inclusion, and the full noise analysis are all redone at the IPTA level, largely (checked) by hand. For
completeness and cross-checking, the whole combination is nowadays carried out
twice, with two semi-independent software stacks: one based on
\texttt{PINT}/\texttt{enterprise}, and one based on
\texttt{Tempo2}/\texttt{TempoNest}. This thoroughness has resulted in datasets that have been of great value to the community~\citep{perera2019international,antoniadis2022international}. However, the combination projects have suffered delays of many years, they
place a great strain on the collaboration, and, as discussed in
Sec.~\ref{sec:intro}, individual PTA data releases have scooped unreleased IPTA datasets. Moreover, the duplication across two software
stacks is no longer necessary: as summarized in
Sec.~\ref{sec:roundtripconsistency}, a fully specified timing model is evaluated by \texttt{PINT} and \texttt{Tempo2} with differences of only a few nanoseconds, so the two stacks are equivalent where it matters.

The way forward, we believe, is to deconstruct the combination effort into its
four distinct tasks:
\begin{enumerate}
    \item TOA generation from the pulse profiles and initial timing model definition;
    \item the actual data combination;
    \item timing model refinement and parameter inclusion;
    \item noise modeling;
\end{enumerate}
and to recognize that these are separate responsibilities that need not, and
should not, live in a single project.

TOA generation is the responsibility of the individual PTAs, who already carry
it out for their own data releases. When a problem with an individual dataset
surfaces at the combination stage, we should not patch the combined dataset. Instead, the issue should be reported back to the PTA, like a bug report, and
the PTA should fix it in an updated release. The data combination itself, as
this paper demonstrates, is mechanical. \texttt{MetaPulsar} constructs a
consistent combined timing model in under two hours
(Sec.~\ref{sec:iptadr2gwb}), with no manual curation of
\texttt{.par} files.

We argue that timing model refinement should be done after the combination. With the
longer time span and improved sensitivity of a combined dataset, parameters
that were not significant in any individual dataset can become measurable, and
the traditional response is to add them to the model one significance test at a
time. We advocate the simpler and physically correct alternative of
Sec.~\ref{sec:marginalization}: include the full model, with all parameters
that describe the system, and place proper priors on them. The posterior of a poorly determined parameter is then simply equal to its prior. There is no need to leave such a parameter out of the model, and the marginalization remains well
behaved. While we consider this approach out of scope for this paper, we note
that it is a simple matter to implement in existing PTA analysis codes. For
instance, it is straightforward to modify \texttt{MetaPulsar} to replace the
timing model of the reference PTA with a user-provided timing model.

Finally, noise modeling is not needed for data combination. Typically a
full noise analysis is carried out before a
combined \texttt{.par} file is considered valid. As pointed out in
Sec.~\ref{sec:deterministicsum}, noise parameters do not enter the delay
calculation, so they cannot affect the combination. In practice the noise
analysis is redone anyway by any downstream analysis: the initial data release
papers ship with simple noise models, and the advanced noise modeling arrives
years later, together with the GW searches that are built on it.

With the tasks separated, the IPTA workflow we propose is the following.
\begin{enumerate}
    \item The IPTA does not produce combined datasets itself. Analyses ingest
      the individual PTA data releases directly, combined on the fly by tools
      such as \texttt{MetaPulsar}.
    \item The IPTA focuses its combination expertise on data quality and
      cross-PTA consistency checks: quantitative tests of whether the datasets
      of a pulsar can be validly combined. The checks include iterating the
      timing-model expansion to the generalized-least-squares maximum and
      comparing the solution at the two expansion points, which flags the
      parameters for which the linearization fails. The broader program of
      checks will be discussed in a future follow-up paper in this series.
    \item Every time a PTA publishes a data release, the IPTA runs a
      rapid view of the updated combination. Since the combination is
      automated, this can be fully scripted, and given the timings of
      Sec.~\ref{sec:iptadr2gwb}, even a full analysis can be turned around within a
      week. The results are posted publicly, on a project page or on the
      arXiv.
    \item The flagship IPTA projects concentrate on advanced noise modeling
      and on the GW analyses that are built on top of it.
\end{enumerate}

This division of labor also addresses the concern of being scooped. The rapid views are straightforward, and they cannot be scooped when the IPTA publishes them immediately. The flagship analyses depend on expertise that is concentrated in the IPTA collaborations themselves. What remains is the
full IPTA data release project, and we believe it is possible to restructure the effort after IPTA
DR3. We believe a more targeted approach based on data quality and
consistency checks prior to release is complete and efficient, and will allow faster science results with less delay between data-taking and publication.

\section{Single-pulsar tests on released PTA data}
\label{sec:singlepulsartests}

\begin{figure*}
\includegraphics[width=1.0\linewidth]{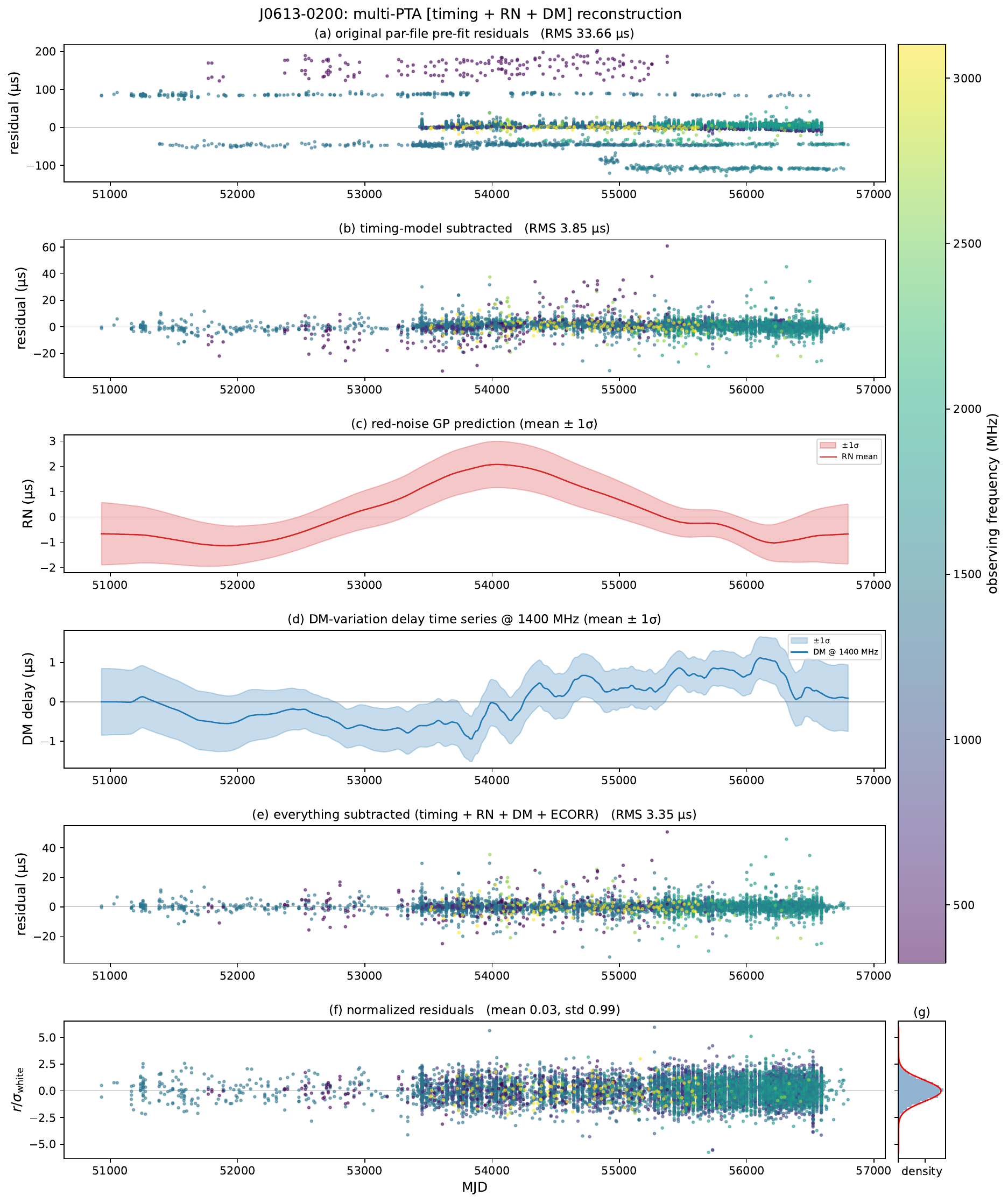}
\caption{Residual decomposition for the shared \texttt{MetaPulsar}
combination of PSR~J0613$-$0200, with all noise hyperparameters held at
their empirical-Bayes values. Panel (a) shows the pre-fit timing
residuals of the combined dataset. Panel (b) shows the residuals after
subtraction of the maximum-posterior timing model. Panels (c) and (d) show
the reconstructed red-noise and dispersion-measure waveforms with their
$1\sigma$ bands. Panel (e) shows the residuals after subtraction of the
timing model, both waveforms, and the epoch-correlated \texttt{ECORR}
terms, which should be white. Panel (f) shows
those residuals normalized by their per-TOA uncertainties, which should
follow a univariate Gaussian, and panel (g) confirms this with a histogram
of the normalized residuals against the standard normal distribution.}
\label{fig:j0613_waveforms}
\end{figure*}

\begin{figure*}
\includegraphics[width=1.0\linewidth]{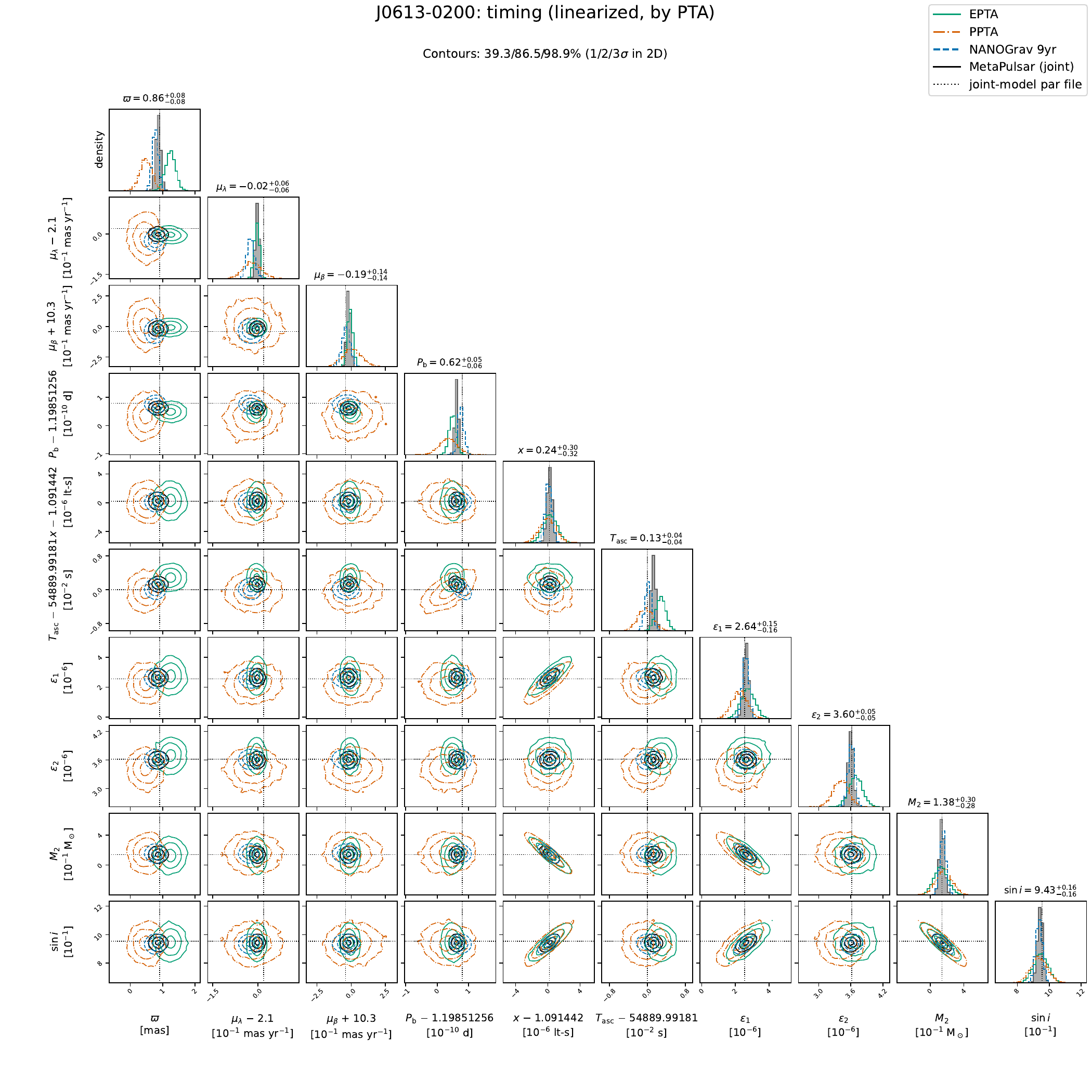}
\caption{Linearized timing posteriors for PSR~J0613$-$0200: the
individual EPTA DR1, PPTA DR2, and NANOGrav 9-yr datasets, each analyzed on
its own, overplotted with the shared \texttt{MetaPulsar} combination
(filled). All four analyses expand the same shared timing model, the one
\texttt{MetaPulsar} writes for the combination, whose parameter values are
marked by the dotted lines, so the posteriors are directly comparable. The
timing model is marginalized as in Sec.~\ref{sec:marginalization} and the
timing offsets are drawn from their Gaussian conditional for every retained
sample of the white-noise, red-noise, and dispersion-measure hyperparameters; the values
above the diagonal panels are the medians and $16$--$84\%$ ranges of the
combination. The noise model is Eq.~\eqref{eq:flattail}, and the white noise is sampled
under the priors of Table~\ref{tab:noisepriors}. The combined posterior is the narrowest and lies within the region the individual datasets support, except for the
parallax (Sec.~\ref{sec:consistency}). The parameters shown are the parallax
$\varpi$, the ecliptic proper motions $\mu_\lambda$ and $\mu_\beta$, and the
binary parameters $P_b$, $x$, $T_{\rm asc}$, $\varepsilon_1$,
$\varepsilon_2$, $M_2$, and $\sin i$ (Table~\ref{tab:binaryparams}). The
$M_2$ and $\sin i$ panels are not a measurement of those parameters: all
four posteriors are linearized at the same point, so the comparison between
datasets is valid.}
\label{fig:j0613_pta_corner}
\end{figure*}

In this Section we test the methods of Secs.~\ref{sec:theory} and~\ref{sec:metapulsar} on released PTA data, one pulsar at a time. The datasets are the
component of IPTA DR2, assembled as a mixed \texttt{PINT}+\texttt{Tempo2}
combination, the same combination used for later releases. EPTA DR1 and PPTA
DR2 are processed with \texttt{Tempo2}. The NANOGrav 9-yr release requires
\texttt{TEMPO}, which \texttt{MetaPulsar} does not support, so that dataset is rewritten onto the common convention profile of
Sec.~\ref{sec:fullspecification} and processed with \texttt{PINT}.
Sec.~\ref{sec:roundtripconsistency} shows that \texttt{PINT} and
\texttt{Tempo2} agree on those rewritten files. For each
pulsar we build the shared \texttt{MetaPulsar} combination of
Sec.~\ref{sec:metapulsar}, and we analyze all datasets with one and the same
model. The white noise uses \texttt{EFAC} and \texttt{EQUAD} per backend group, and
epoch-correlated \texttt{ECORR} terms wherever a backend records several TOAs
per observation. On top of the white noise we include a red-noise GP and a
chromatic dispersion-measure GP, both of the form of Eq.~\eqref{eq:flattail}, on the
Fourier grid of Sec.~\ref{sec:iptadr2gwb} ($47$ and $159$ components). The
dispersion-measure delay scales as $\nu^{-2}$ with radio frequency $\nu$.
The empirical-Bayes maximum is the set of noise hyperparameters that jointly maximize the single-pulsar marginal likelihood of Eq.~\eqref{eq:marginal} of that pulsar's combination. The generalized-least-squares (GLS) maximum is the timing model at that same point. Whenever noise parameters are held fixed in this paper, they are fixed at these values. Figs.~\ref{fig:j0613_waveforms}
and~\ref{fig:j0613_triple} hold every noise hyperparameter there.
Fig.~\ref{fig:j0613_pta_corner} samples the white-noise, red-noise, and dispersion-measure hyperparameters under the priors of Table~\ref{tab:noisepriors}.
The timing model is treated in
the fully linearized limit of Sec.~\ref{sec:marginalization}: all timing
parameters are marginalized analytically. Posteriors are obtained with the
\texttt{nltiming} package~\citep{nltiming}.

\subsection{Test 1: does the model explain the data}
\label{sec:modelexplaindata}

As a first test we check whether the combined model describes the combined data. In Fig.~\ref{fig:j0613_waveforms} we show the residuals of PSR~J0613$-$0200 after subtracting the components of the model one by one. The
pre-fit residuals of panel (a) show large structure, which is expected: as
argued in Sec.~\ref{sec:timingmodelvalues}, the parameter values in the
released \texttt{.par} files carry no special meaning for the combined
dataset, and they often lie several standard deviations away from the joint posterior maximum---especially if the timing model was timescale-converted from TCB to TDB. Subtracting the maximum-posterior timing model (panel b)
removes that structure and leaves the stochastic contributions, which the
reconstructed red-noise and dispersion-measure waveforms (panels c and d)
then absorb. Panel (e) also subtracts the \texttt{ECORR} terms, which are
correlated between the TOAs of one observation and would otherwise not be
white; for this pulsar they amount to $30\,\mathrm{ns}$ RMS. What
remains is consistent with white noise, and the
normalized residuals (panels f and g) are well described by a unit Gaussian. We conclude that the combined model, built mechanically from the released \texttt{.par} and \texttt{.tim} files, describes the combined data.

\subsection{Test 2: cross-PTA consistency}
\label{sec:crossptaconsistency}

The second test is one of cross-PTA consistency. Figure~\ref{fig:j0613_pta_corner}
shows the posteriors of the timing parameters of
PSR~J0613$-$0200, obtained separately for each of the three individual
datasets, together with the posterior of the shared combination. We note that the three single-PTA analyses do not use the released
\texttt{.par} files. They use the per-PTA \texttt{.par} files that
\texttt{MetaPulsar} writes for the combination, in which the shared
parameters carry the agreed values and the binary model is the family of
the combination, so that all four posteriors describe the same parameters
expanded about the same point. Without that step, the four posteriors of
the projected semi-major axis \texttt{A1} (Table~\ref{tab:binaryparams})
would not be the same quantity in the four analyses.
The orbital and proper-motion posteriors agree with one another, which is
the underlying requirement for any valid data combination, and the combined
posterior lies within the region they jointly support, with the tighter
constraints one expects from combining information. The parallax does not
agree: EPTA DR1 and PPTA DR2 differ by $0.76\pm0.24\,\mathrm{mas}$.
Sec.~\ref{sec:consistency} repeats the comparison for every pulsar observed
by more than one PTA, for this power law and for a free spectrum. This
figure is the data quality check we advocate in
Sec.~\ref{sec:proposedworkflow}: an inconsistent parameter flags the pulsar
for a report back to the PTAs.

\begin{figure*}
\includegraphics[width=1.0\linewidth]{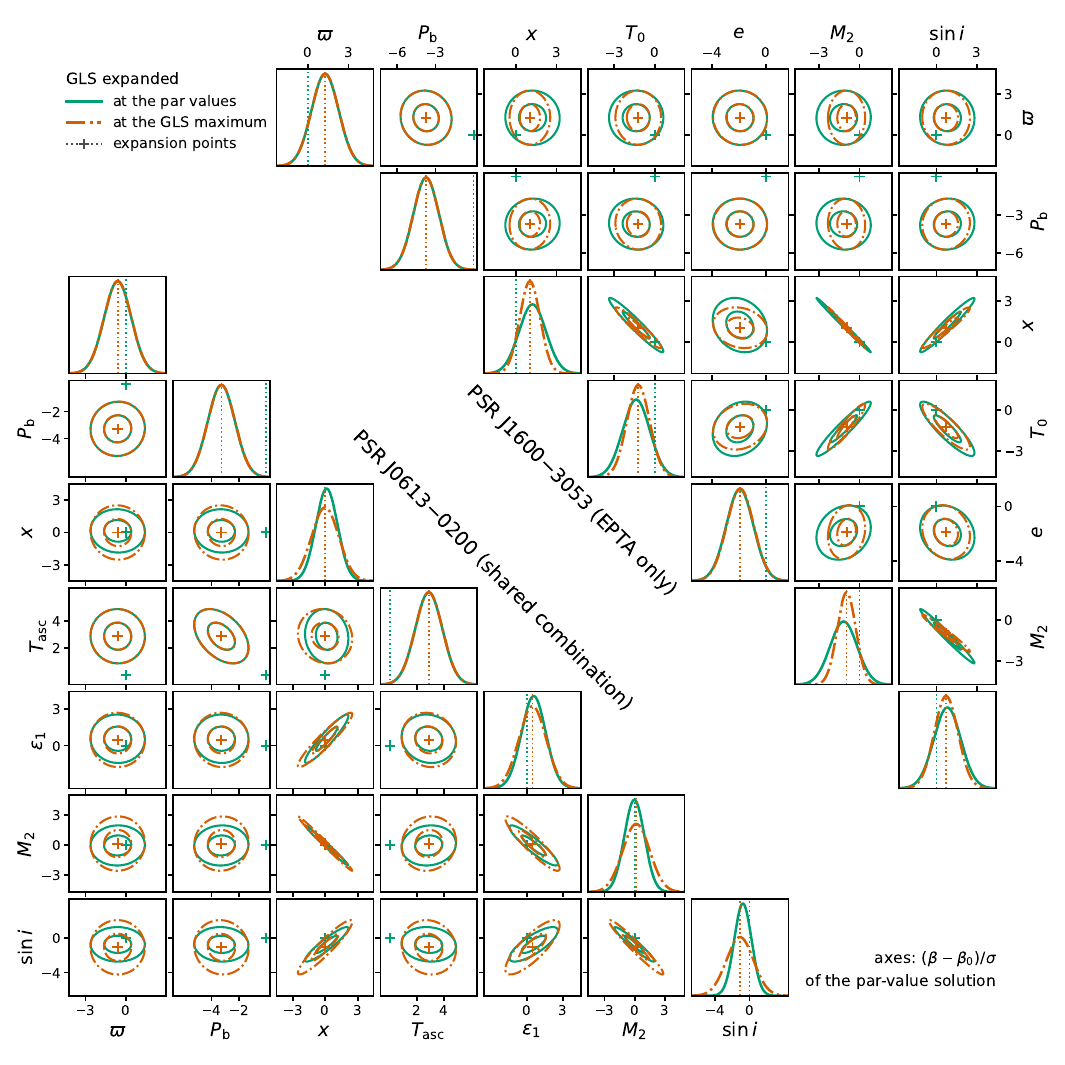}
\caption{Generalized-least-squares timing solutions for the shared
combination of PSR~J0613$-$0200 (lower left) and the EPTA DR1 data of
PSR~J1600$-$3053 (upper right), at the empirical-Bayes noise hyperparameters. Each solution is
evaluated at two expansion points: the released \texttt{.par}-file values, and
the GLS maximum of that dataset, reached by re-expanding at each
solution until the solution coincides with its expansion point. The contours
are the $1\sigma$ and $2\sigma$ ellipses of the two Gaussian solutions, and
the crosses and dotted lines mark the two expansion points. Every axis shows
the offset from the \texttt{.par}-file value, in units of the standard
deviation of the solution expanded there. The two solutions coincide for
every parameter except those constrained by the Shapiro delay: the companion
mass $M_2$, $\sin i$, and the projected semi-major axis $x$, together
with $\varepsilon_1$ for PSR~J0613$-$0200 and the epoch of periastron $T_0$ for
PSR~J1600$-$3053.}
\label{fig:j0613_triple}
\end{figure*}

\subsection{Test 3: initial-value dependence}
\label{sec:initialvalue}

The third test checks whether the timing-model parameters depend on the
initial values in the \texttt{.par} files. We do not need a full MCMC over the noise hyperparameters for this test. Conditional on the
hyperparameters, the offsets $\bm{\varepsilon}$ of Eq.~\eqref{eq:linear} are
Gaussian, and the mean of that conditional is the generalized-least-squares
solution. If this conditional does not depend on the expansion point, the
same is true at every other choice of hyperparameters. Checking the GLS
solution is therefore sufficient, and it is more accurate and faster than
sampling.

Figure~\ref{fig:j0613_triple} shows the check for the shared combination of
PSR~J0613$-$0200 and for the EPTA DR1 data of PSR~J1600$-$3053. For each pulsar we evaluate the GLS
solution at the empirical-Bayes hyperparameters, timing model once expanded at the
released \texttt{.par} values and once re-expanded when iterated to the GLS
maximum. The second expansion re-evaluates the full timing model at the new parameter values, and no new \texttt{.par} file is written. The \texttt{.par} values lie many $\sigma$ away from the GLS solution, and
moving the expansion point there changes the result only mildly. The centers
of the posteriors move by a fraction of a standard deviation, and only the
parameters tied to the Shapiro delay change at all. On
PSR~J0613$-$0200 the widths of $\sin i$, $M_2$, $x$, and $\varepsilon_1$
become $57\%$, $35\%$, $25\%$, and $12\%$ larger, while their centers move by
$0.34\sigma$, $0.18\sigma$, $0.15\sigma$, and $0.10\sigma$. On the EPTA data
of PSR~J1600$-$3053 the widths of $M_2$, $x$, $T_0$, and $\sin i$ become
$32\%$, $25\%$, $16\%$, and $12\%$ smaller, and their centers move by
$0.22\sigma$, $0.19\sigma$, $0.15\sigma$, and $0.09\sigma$. Every other
parameter in Fig.~\ref{fig:j0613_triple} is unchanged, to a percent in the
width and to a hundredth of a standard deviation in the center. The red-noise and dispersion-measure posteriors of these
two pulsars are consistent between the two expansion points. Across the $57$ PTA--pulsar datasets of the pulsars observed by more than one PTA, the free-spectrum noise posteriors from the linearized timing model and from a nonlinear analysis with \texttt{nltiming} are consistent as well. That nonlinear analysis is the subject of a later paper in this series. The median of the largest shift in any frequency bin is $0.1\sigma$, and one dataset, PSR~J2317$+$1439 in NANOGrav, has a red-noise bin at $1.1\sigma$. We conclude that the datasets are quite robust against modifications of the values of the \texttt{.par} file.

The differences are limited to the Shapiro-delay parameters. Where the posterior of a
binary parameter is manifestly non-Gaussian, or where a parameter is so poorly
determined that a physical prior, the non-negativity of a parallax for
instance, visibly shapes its posterior, the linear treatment of
Sec.~\ref{sec:marginalization} reaches its limits and the implicit unbounded
Gaussian prior of Eq.~\eqref{eq:marginal} extends into unphysical territory.
This is not caused by the direct combination. The same considerations regarding a linear versus a nonlinear treatment apply to a manual data combination, and current PTA analyses already deal with them. In this paper \texttt{nltiming} is used only to evaluate the linearized likelihood. The fully nonlinear treatment of the timing model, which \texttt{nltiming} also provides, is the subject of a forthcoming paper.

We mention one issue with the public data here, because it does not show up in an analysis with a single timing package. It only becomes visible when a second timing package reads the same files. The EPTA DR1 source tree of PSR~J1939$+$2134 contains
early Nan\c{c}ay DDS observations (MJD 47958--51474) whose observatory clock
chain in \texttt{PINT} runs through a clock file that begins only at
MJD~52974. \texttt{PINT} extrapolates the missing clock correction, \texttt{Tempo2} does not, and the two residual series disagree at the $65\,\mathrm{ns}$ level on
those TOAs. Those early TOAs are not part of the released dataset, but an
earlier version of our processing that ingested them measured this discrepancy. This shows that the clock-correction realization is a data-release responsibility in the sense of
Sec.~\ref{sec:proposedworkflow}. Separately, and for data quality reasons, we exclude the EPTA dataset of
PSR~J1939$+$2134 from the combination altogether ($991$ TOAs), so
that this pulsar enters through its NANOGrav 9-yr and PPTA DR2 data only.
This is in line with the decision of the IPTA in subsequent releases.

A second type of issue becomes visible only when the noise model is fitted, and we exclude two pieces of data because of it. Our empirical-Bayes white-noise
fit (Sec.~\ref{sec:iptadr2gwb}) bounds each backend's \texttt{EFAC} to $[0.1,10]$.
Across all pulsars of the combination, three backends drive \texttt{EFAC} to the
\emph{upper} bound: PSR~J1738$+$0333 Jodrell \texttt{DFB.1520}
($56$ TOAs, $770\,\mu\mathrm{s}$ residual RMS), and PSR~J2019$+$2425
Nan\c{c}ay \texttt{BON.1400} and \texttt{BON.1600} ($44$ and $27$ TOAs,
$1106$ and $722\,\mu\mathrm{s}$). Every other \texttt{EFAC} that is at a bound is at the \emph{lower} one, which is the familiar \texttt{EFAC}--\texttt{EQUAD} degeneracy and is harmless. The upper bound is a problem. Because the GWB analysis holds white noise fixed at this fit, a backend that needs more noise than the bound allows enters
the likelihood with its scatter under-modelled, and its residuals are therefore
over-weighted. Raising the bound does not help, since an \texttt{EFAC} of many tens means that the quoted TOA uncertainties carry no information.

We treat the two cases differently. For
PSR~J1738$+$0333 the defect is confined to one backend: its two Nan\c{c}ay
backends ($199$ and $63$ TOAs, $10.0$ and $8.4\,\mu\mathrm{s}$ RMS, \texttt{EFAC}
$1.14$) and its four NANOGrav backends ($4.8$--$9.4\,\mu\mathrm{s}$, \texttt{EFAC}
$0.94$--$1.15$) are fine, so we disable the Jodrell \texttt{INCLUDE} alone and
retain $2885$ of $2941$ TOAs, keeping the pulsar in both PTAs. For
PSR~J2019$+$2425 both of these backends are Nan\c{c}ay, and removing them
leaves $59$ Jodrell TOAs; as the pulsar is observed by EPTA alone, too little
remains to constrain a nanohertz signal, and we drop its EPTA dataset entirely,
which removes the pulsar from the combination. Both edits are recorded as
commits on our fork of the IPTA DR2 source tree, so that the exclusions are part of the dataset, and not of the analysis settings. After the two exclusions and the
dropped EPTA dataset of PSR~J1939$+$2134, the combination contains $62$ pulsars and
$199\,934$ TOAs over $18.32\,\mathrm{yr}$, assembled from $95$ PTA--pulsar
datasets, one per pulsar per PTA.

\section{Cross-PTA consistency of the timing parameters}
\label{sec:consistency}

The corner plot in Figure~\ref{fig:j0613_pta_corner} concerns one pulsar. The same question, whether
the datasets of the individual PTAs measure the same timing model, can be
asked of every pulsar that more than one collaboration observes, and it is
the data-quality check of Sec.~\ref{sec:proposedworkflow} that we believe the
IPTA should run at every release. Of the $62$ pulsars in the combination,
$24$ are observed by two or three of the PTAs, and $9$ by all three. For
each of them we analyze the individual datasets separately on the shared
timing model of the combination, and we compare the marginal posteriors of
every shared timing parameter between each pair of PTAs. The timing model is
fully linearized and marginalized, as in Sec.~\ref{sec:marginalization}, and
the timing offsets are drawn from their Gaussian conditional. The noise model differs from that of Sec.~\ref{sec:singlepulsartests}. White noise is sampled, under
the priors of Table~\ref{tab:noisepriors}, and the red-noise and
dispersion-measure processes are free spectra, one amplitude per Fourier bin
on each dataset's own span, down to periods of $20$ and $6$ weeks. Every single-PTA analysis expands the same
per-PTA \texttt{.par} file that \texttt{MetaPulsar} writes for the shared
combination, so the parameter sets of the single-PTA datasets of one pulsar are identical.

We quantify the comparison with the two-sample tension statistic of
\citet{jacobTensionStatistic2026}, developed there to compare PTA and electromagnetic
constraints on the same quantity. For two posteriors $\mathcal{P}_1(\theta)$
and $\mathcal{P}_2(\theta)$ of the same scalar parameter, with the same prior
support, the difference $\delta = \theta_2 - \theta_1$ has density
$\mathcal{P}_\Delta(\delta) = \int \mathcal{P}_2(\delta+\theta)\,
\mathcal{P}_1(\theta)\,\mathrm{d}\theta$, and the two-sided tension
\begin{equation}
\mathcal{T}_2 \;=\; \int_{\mathcal{P}_\Delta(\delta)\,\geq\,\mathcal{P}_\Delta(0)}
\mathcal{P}_\Delta(\delta)\,\mathrm{d}\delta
\label{eq:tension}
\end{equation}
is the highest-posterior-density credible level at which ``no shift'',
$\delta=0$, is excluded~\citep[after][]{raveriConcordanceDiscordance2019}.
For two Gaussians it reduces to
$\mathcal{T}_2 = 2\Phi\!\left(|\mu_2-\mu_1|/\sqrt{\sigma_1^2+\sigma_2^2}\right)-1$,
so that $\mathcal{T}_2 = 68\%$ corresponds to a $1\sigma$ shift and $95\%$ to
$2\sigma$. We note that a low $\mathcal{T}_2$ does not confirm agreement. It only means that there is insufficient evidence for a shift. Following \citet{jacobTensionStatistic2026} we reject consistency at $\mathcal{T}_2 > 95\%$. The statistic is evaluated on
the MCMC draws directly, by histogramming both chains and convolving them. The statistic is one-dimensional and ignores the covariances between parameters, so it can over- or understate a tension in a direction that is correlated between parameters.

We omit $\mathrm{DM}$, $F_0$, $F_1$, $\mathrm{DM}1$, and $\mathrm{DM}2$
from the comparison in Table~\ref{tab:tension}. Those spin and
dispersion-measure polynomial terms are degenerate with the lowest modes of
the red-noise and dispersion-measure processes, and each single-PTA dataset has its own data span, so the marginal posteriors leak through the window function and
are not a cross-PTA test~\citep{crisostomiDiagonalApproximationsImproved2025}.

\begin{table*}
\caption{Two-sided tension $\mathcal{T}_2$ of Eq.~\eqref{eq:tension}, in
percent, between the single-PTA posteriors of every pulsar observed by more
than one PTA (E: EPTA DR1, N: NANOGrav 9-yr, P: PPTA DR2). Each single-PTA
analysis is the linearized, timing-marginalized analysis of
Sec.~\ref{sec:marginalization}, with free-spectrum red-noise and
dispersion-measure processes (Sec.~\ref{sec:consistency}), on the per-PTA
\texttt{.par} file of the shared combination. This page gives the
astrometric parameters; the binary parameters follow. Pairs with
$\mathcal{T}_2 > 95\%$ are set in bold.}
\label{tab:tension}
\centering
\begin{tabular}{llrrrrr}
\hline
Pulsar & PTAs & \texttt{ELONG} & \texttt{ELAT} & \texttt{PMELONG} & \texttt{PMELAT} & \texttt{PX} \\
\hline
B1855$+$09 & E--N & 59 & 40 & 17 & 2 & 51 \\
 & E--P & 29 & 0 & 88 & 25 & 3 \\
 & N--P & 56 & 37 & 81 & 25 & 56 \\
B1937$+$21 & N--P & 26 & 14 & 45 & 4 & 8 \\
B1953$+$29 & E--N & 54 & 70 & 48 & 76 & 25 \\
J0030$+$0451 & E--N & 29 & 43 & 7 & 27 & 45 \\
J0613$-$0200 & E--N & 89 & 13 & 40 & 63 & 79 \\
 & E--P & 0 & 34 & 17 & 26 & 84 \\
 & N--P & 89 & 61 & 32 & 67 & 66 \\
J1012$+$5307 & E--N & 11 & 60 & 63 & 42 & 30 \\
J1022$+$1001 & E--P & 8 & 22 & 87 & 82 & 53 \\
J1024$-$0719 & E--N & 0 & 34 & 10 & 21 & 4 \\
 & E--P & 41 & 7 & 37 & 15 & 46 \\
 & N--P & 0 & 33 & 30 & 39 & 60 \\
J1455$-$3330 & E--N & 77 & 73 & 53 & 21 & 57 \\
J1600$-$3053 & E--N & 31 & 17 & 27 & 65 & 0 \\
 & E--P & 28 & 15 & 9 & 27 & 11 \\
 & N--P & 23 & 16 & 19 & 80 & 16 \\
J1640$+$2224 & E--N & 31 & 31 & 26 & 25 & 26 \\
J1643$-$1224 & E--N & 15 & 0 & 71 & 59 & 38 \\
 & E--P & 61 & 43 & 26 & 86 & 8 \\
 & N--P & 12 & 0 & 46 & 88 & 28 \\
J1713$+$0747 & E--N & 58 & 87 & 51 & 25 & 16 \\
 & E--P & 25 & 40 & 52 & 13 & 56 \\
 & N--P & 53 & 83 & 2 & 12 & 63 \\
J1730$-$2304 & E--P & 0 & 0 & 40 & 44 & 62 \\
J1738$+$0333 & E--N & 74 & 67 & 54 & 70 & 39 \\
J1744$-$1134 & E--N & 49 & 36 & 82 & 68 & 38 \\
 & E--P & 61 & 66 & 67 & 68 & 87 \\
 & N--P & 51 & 26 & 56 & 85 & \textbf{96} \\
J1853$+$1303 & E--N & 41 & 7 & 2 & 4 & 29 \\
J1909$-$3744 & E--N & 25 & 27 & 15 & 24 & 89 \\
 & E--P & 53 & 78 & 20 & 16 & 66 \\
 & N--P & 26 & 27 & 0 & 44 & 71 \\
J1910$+$1256 & E--N & 28 & 4 & 2 & 61 & 61 \\
J1918$-$0642 & E--N & 36 & 0 & 21 & 24 & 29 \\
J2010$-$1323 & E--N & 7 & 44 & 14 & 48 & 70 \\
J2124$-$3358 & E--P & 5 & 35 & 49 & 36 & 37 \\
J2145$-$0750 & E--N & 41 & 46 & 15 & 76 & 0 \\
 & E--P & 31 & 22 & 17 & 49 & 30 \\
 & N--P & 34 & 44 & 10 & 78 & 26 \\
J2317$+$1439 & E--N & 15 & 9 & 7 & 17 & 86 \\
\hline
\end{tabular}

\end{table*}

\begin{table*}
\contcaption{Binary parameters, grouped by model. The upper block is the
ELL1 family and the lower block is DD, including the orthometric and DDK models. A dash marks a parameter absent from that pulsar's model. The $M_2$
and $\sin i$ columns are the linearized comparison.}
\centering
\begin{center}
\begin{tabular}{llrrrrrrrrrrr}
\multicolumn{13}{c}{ELL1} \\
\hline
Pulsar & PTAs & \texttt{PB} & \texttt{PBDOT} & \texttt{A1} & \texttt{A1DOT} & \texttt{TASC} & \texttt{EPS1} & \texttt{EPS2} & \texttt{M2} & \texttt{SINI} & \texttt{EPS1DOT} & \texttt{EPS2DOT} \\
\hline
J0613$-$0200 & E--N & 83 & -- & 6 & -- & -- & 6 & 1 & 4 & 8 & -- & -- \\
 & E--P & 76 & -- & 16 & -- & -- & 28 & 72 & 14 & 4 & -- & -- \\
 & N--P & 94 & -- & 24 & -- & -- & 20 & 80 & 11 & 19 & -- & -- \\
J1012$+$5307 & E--N & \textbf{100} & -- & \textbf{99} & -- & -- & 1 & 2 & -- & -- & -- & -- \\
J1738$+$0333 & E--N & 21 & -- & 16 & -- & 6 & \textbf{96} & 71 & -- & -- & -- & -- \\
J1909$-$3744 & E--N & 67 & 53 & 12 & -- & 0 & 2 & 83 & 3 & 84 & -- & -- \\
 & E--P & 4 & 38 & 86 & -- & 0 & 41 & 91 & 51 & 95 & -- & -- \\
 & N--P & 77 & 82 & 84 & -- & -- & 34 & 4 & 29 & 49 & -- & -- \\
J2317$+$1439 & E--N & 59 & -- & 95 & 85 & -- & \textbf{98} & 37 & -- & -- & 84 & \textbf{100} \\
\hline
\end{tabular}
\vspace{6pt}
\begin{tabular}{llrrrrrrrrrrrrrr}
\multicolumn{16}{c}{DD, DDH, and DDK} \\
\hline
Pulsar & PTAs & \texttt{PB} & \texttt{PBDOT} & \texttt{A1} & \texttt{A1DOT} & \texttt{T0} & \texttt{ECC} & \texttt{OM} & \texttt{OMDOT} & \texttt{M2} & \texttt{SINI} & \texttt{H3} & \texttt{KIN} & \texttt{KOM} & \texttt{STIGMA} \\
\hline
B1855$+$09 & E--N & 92 & -- & 93 & -- & 2 & 91 & 57 & -- & 91 & 71 & -- & -- & -- & -- \\
 & E--P & 80 & -- & 37 & -- & 13 & 49 & 67 & -- & 25 & 88 & -- & -- & -- & -- \\
 & N--P & 28 & -- & 73 & -- & 12 & \textbf{98} & 34 & -- & 92 & \textbf{96} & -- & -- & -- & -- \\
B1953$+$29 & E--N & 81 & -- & 43 & 84 & 28 & 21 & 91 & -- & -- & -- & -- & -- & -- & -- \\
J1022$+$1001 & E--P & 6 & 28 & 29 & \textbf{98} & 9 & 34 & 6 & 10 & -- & -- & 21 & -- & -- & 45 \\
J1455$-$3330 & E--N & 80 & -- & 33 & 15 & -- & \textbf{98} & 18 & -- & -- & -- & -- & -- & -- & -- \\
J1600$-$3053 & E--N & \textbf{98} & -- & 67 & -- & -- & 45 & 72 & \textbf{98} & 61 & 35 & -- & -- & -- & -- \\
 & E--P & 50 & -- & 28 & -- & -- & 77 & 56 & 49 & 1 & 3 & -- & -- & -- & -- \\
 & N--P & \textbf{99} & -- & 74 & -- & 8 & 85 & 91 & \textbf{99} & 58 & 32 & -- & -- & -- & -- \\
J1640$+$2224 & E--N & 58 & -- & 52 & 55 & 0 & 39 & 49 & 58 & 51 & 38 & -- & -- & -- & -- \\
J1643$-$1224 & E--N & 86 & -- & 26 & 34 & -- & 25 & 48 & -- & -- & -- & -- & -- & -- & -- \\
 & E--P & 81 & -- & 42 & 0 & -- & 37 & 18 & -- & -- & -- & -- & -- & -- & -- \\
 & N--P & 30 & -- & 45 & 17 & -- & 19 & 70 & -- & -- & -- & -- & -- & -- & -- \\
J1713$+$0747 & E--N & 46 & -- & 17 & -- & 10 & 78 & 38 & -- & 28 & -- & -- & 28 & 26 & -- \\
 & E--P & 78 & -- & 55 & -- & 0 & 66 & 55 & -- & 65 & -- & -- & 55 & 58 & -- \\
 & N--P & 58 & -- & 58 & -- & 12 & 4 & 41 & -- & 59 & -- & -- & 42 & 56 & -- \\
J1853$+$1303 & E--N & 4 & -- & 2 & 11 & 10 & 7 & 60 & -- & -- & -- & -- & -- & -- & -- \\
J1910$+$1256 & E--N & 3 & -- & 59 & 40 & -- & 16 & 74 & -- & -- & -- & -- & -- & -- & -- \\
J1918$-$0642 & E--N & 45 & -- & 14 & -- & -- & 38 & 46 & -- & 33 & 48 & -- & -- & -- & -- \\
J2145$-$0750 & E--N & \textbf{97} & -- & 18 & 86 & -- & 38 & 78 & -- & -- & -- & -- & -- & -- & -- \\
 & E--P & 89 & -- & \textbf{98} & 10 & -- & 10 & 75 & -- & -- & -- & -- & -- & -- & -- \\
 & N--P & 7 & -- & 91 & 76 & 2 & 41 & \textbf{97} & -- & -- & -- & -- & -- & -- & -- \\
\hline
\end{tabular}
\end{center}

\end{table*}

Table~\ref{tab:tension} reports $\mathcal{T}_2$ for the \tensionNpulsars\ pulsars observed by more than one PTA: \tensionNpairs\ (pulsar, PTA-pair) combinations and \tensionNtriples\ (pulsar, pair, parameter) comparisons. Of these, \tensionNflagged\ (\tensionFracFlagged\%) exceed the $95\%$ threshold and are set in bold. Sixteen of the seventeen flags are binary parameters. The only astrometric flag is the NANOGrav--PPTA parallax of PSR~J1744$-$1134, one of \tensionNpx\ parallax comparisons. The $M_2$ and $\sin i$ columns are the linearized comparison. The NANOGrav--PPTA $\sin i$ entry for PSR~B1855$+$09, at $96\%$, is confirmed by the nonlinear analysis with \texttt{nltiming} described in Sec.~\ref{sec:singlepulsartests} and presented in a later paper, which flags the same pair at $99.9\%$ in $\sin i$ and $98.2\%$ in $M_2$.

We make two remarks. First, the response of the timing residuals to the astrometric parameters contains an annual and a semi-annual frequency. Specifically, the parallax signature has a period of half a year, so the noise model has to reach at least those frequencies. While we construct the power-law model of Eq.~\eqref{eq:flattail} (flat tail included) to do so, one amplitude, one index, and a floor are too rigid there to follow a parallax-like mismatch. Table~\ref{tab:tension} therefore uses a free spectrum for both the red noise and the dispersion-measure variations, one amplitude per Fourier bin for each, which supplies that flexibility and is the more flexible noise model in general. The two treatments give different answers for
PSR~J0613$-$0200: as can be seen in Fig.~\ref{fig:j0613_pta_corner} under the power law the EPTA--PPTA and EPTA--NANOGrav
parallaxes disagree at $\mathcal{T}_2 = 99.8\%$ and $98.4\%$, whereas with
the free spectra the same comparisons give $84\%$ and $79\%$.  Second, a tension does not tell us which of the two datasets is wrong, or whether the model is. It should be reported back to the PTAs, as we advocate in Sec.~\ref{sec:proposedworkflow}, and the combination of a flagged pulsar should be treated with caution until
the cause is understood.

\section{IPTA DR2 GWB analysis comparison}
\label{sec:iptadr2gwb}

As a final test we carry out a GWB search, which is the main purpose of an IPTA data combination. We assemble the
three constituents of IPTA DR2, the EPTA DR1, NANOGrav 9-yr, and PPTA DR2
releases, with the shared combination of Sec.~\ref{sec:metapulsar}, and we
search the result for a gravitational-wave background. The combination
comprises $62$ pulsars; $24$ of them are observed by more than one PTA, and
$9$ by all three. Every analysis below sees the same $199\,934$ TOAs and the
same $18.3\,\mathrm{yr}$ span. We note that our combination is not identical to the IPTA DR2 dataset: the official IPTA DR2 added TOAs and flags on top of the constituent releases, re-derived the timing models, and re-did the noise analysis. We therefore would not expect to reproduce the IPTA DR2 results exactly. Instead, we therefore create our own full combination, which we label ``AEI-DR2'': for each pulsar it is a single \texttt{.par} and \texttt{.tim} file of the combined data that can be read by both \texttt{Tempo2} and \texttt{PINT}.

We search the combination in three ways, all with the same
\texttt{discovery} model. The first is the \texttt{MetaPulsar} combination of Sec.~\ref{sec:metapulsar}, in which each dataset is processed by its own timing package: EPTA and PPTA
are evaluated with \texttt{Tempo2}, and NANOGrav with \texttt{PINT}, as in
Sec.~\ref{sec:singlepulsartests}. The second
and third write the combination out as merged \texttt{.par} and
\texttt{.tim} files (that is our own ``AEI-DR2'' release) and read those files as a single dataset, once with
\texttt{PINT} and once with \texttt{Tempo2}. The three versions differ only in how the residuals and the timing design matrix are produced. The TOAs are identical.

We use a standard analysis. White noise uses \texttt{EFAC} and \texttt{EQUAD} per
backend, with \texttt{ECORR} wherever a backend records several TOAs per observation:
every NANOGrav backend (sub-banded TOAs, $28$ per epoch on average) and the
dual-band CPSR2 $20\,$cm data of PPTA. The EPTA DR1 and remaining PPTA DR2
TOAs are band-averaged, one per observation, so any epoch-correlated jitter
there is absorbed by \texttt{EQUAD}. All white-noise parameters are held fixed at the
empirical-Bayes maximum of the single-pulsar marginal likelihood. Each pulsar carries red noise of the form of Eq.~\eqref{eq:flattail}
on max $47$ Fourier frequencies (down to a period of $20$ weeks) and a
dispersion-measure process of the same form, whose delay scales as $\nu^{-2}$ with radio
frequency $\nu$, on max $159$ frequencies (down to $6$
weeks), with those hyperparameters sampled. The red-noise grid extends beyond
$2\,\mathrm{yr}^{-1}$, so that semi-annual power is modeled by the noise
process rather than absorbed by the astrometric parameters
(Sec.~\ref{sec:consistency}). The dispersion-measure grid extends further,
to capture the shorter-timescale structure of the interstellar medium,
but not beyond $1/(6\,\mathrm{weeks})$: higher frequencies would exceed
the Nyquist frequency of the observing cadence of many pulsars in the
combination. We chose these cutoffs for practical reasons, and we did not carry out a systematic investigation. The timing model is marginalized as in
Sec.~\ref{sec:marginalization}. The common signal is the power-law term of Eq.~\eqref{eq:powerlaw} alone,
with Hellings--Downs spatial correlations on the lowest $14$ frequencies.
Its priors, and those of the per-pulsar red-noise and dispersion-measure
hyperparameters, are the uniform boxes of Table~\ref{tab:noisepriors}. Posteriors are drawn with the No-U-Turn
sampler~\citep{hoffman2014no} in \texttt{discovery}/\texttt{numpyro}.

\begin{figure}
\centering
\includegraphics[width=\linewidth]{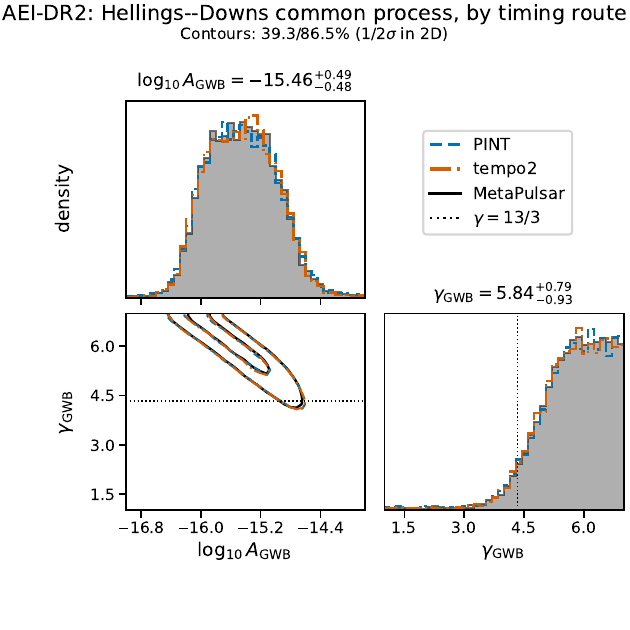}
\caption{Posterior of the amplitude and spectral index of the
Hellings--Downs-correlated common process in the AEI-DR2 combination of
the EPTA DR1, NANOGrav 9-yr, and PPTA DR2 data releases, for three versions of the same data and the same \texttt{discovery} model: the \texttt{MetaPulsar} combination with each dataset processed by its own timing package (filled), and the merged
\texttt{.par}/\texttt{.tim} combination read through \texttt{PINT} alone
(blue dashed) and \texttt{Tempo2} alone (orange dash-dotted). The dotted
line marks the fiducial supermassive-black-hole-binary index
$\gamma=13/3$. The uniform priors are
$\log_{10} A_{\rm GWB}\sim\mathcal{U}(-20,-10)$ and
$\gamma_{\rm GWB}\sim\mathcal{U}(1,7)$.}
\label{fig:gwbcorner}
\end{figure}

The results are shown in Fig.~\ref{fig:gwbcorner}. The three posteriors agree to well within a tenth of the posterior width. This shows that rewriting the combination as a pair of \texttt{.par}/\texttt{.tim} files
and reading it with either timing package recovers the same
Hellings--Downs posterior as the combination in which each dataset is processed by its own timing package. The posterior
is bounded from above but not from below, as IPTA DR2 found for its
Hellings--Downs search. The \texttt{MetaPulsar} combination gives
$\log_{10}A_{\rm GWB} = -15.46^{+0.49}_{-0.48}$ and $\gamma_{\rm GWB} =
5.84^{+0.79}_{-0.93}$ (68\% credible), completely consistent with both the
\texttt{Tempo2} and \texttt{PINT} analyses as a manual combination would yield.  The results agree with \citet{antoniadis2022international}.

The low computational cost is what makes the workflow of Sec.~\ref{sec:proposedworkflow} realistic. Every step from the released files to the posterior of Fig.~\ref{fig:gwbcorner} is scripted:
\begin{center}
\small
\begin{tabular}{@{}>{\raggedright\arraybackslash}p{0.68\columnwidth}r@{}}
\hline
step & wall time \\
\hline
combination of $62$ pulsars (\texttt{MetaPulsar}, one core) & $1.5\,\mathrm{h}$ \\
cross-package parity check, $95$ PTA--pulsar datasets & $0.5\,\mathrm{h}$ \\
noise at the empirical-Bayes maximum, $62$ pulsars (one GPU) & $8\,\mathrm{min}$ \\
GWB sampling, $62$ pulsars (one GPU) & $3.5\,\mathrm{h}$ \\
\hline
\end{tabular}
\end{center}
The first two steps were run on a MacBook Pro with an M1 processor, and the last two steps on a single GPU each. In total it takes less than six hours to go from the released data to a rapid view of the GWB posterior.

\section{Conclusions}
\label{sec:conclusions}

We investigated the combination of the data of multiple PTA collaborations
from first principles, and we introduced \texttt{MetaPulsar}, a software
package that carries out such a combination directly from the released
\texttt{.par} and \texttt{.tim} files. \texttt{MetaPulsar} makes the timing
models of the individual data releases mutually consistent, and compatible
with both \texttt{PINT} and \texttt{Tempo2}, without a manual re-assessment
of the timing model. The two timing packages agree on the rewritten timing
models with a median difference of $1.2\,\mathrm{ns}$.

We validated the method on the constituents of IPTA DR2. We find that the
combined model describes the combined data. For all pulsars we tested,
the linearized timing solution does not depend on the parameter values in the
\texttt{.par} files, with the mild exception of the Shapiro-delay parameters, and
the red-noise and dispersion-measure posteriors are not affected. We compared
the timing parameters measured by the different PTAs for every pulsar that is
observed by more than one PTA, and we advocate such a comparison as a data
quality check. A GWB search on the direct combination gives results that are
consistent with the published IPTA DR2 analysis, and the full procedure takes less than six hours.

The final recommendation of this paper is to restructure the full IPTA data
release project after IPTA DR3. Instead, the released datasets of the
individual PTAs can be combined directly with tools such as
\texttt{MetaPulsar}, so that the efforts of the IPTA can be focused on data
quality checks, noise modeling, and GW analyses.

\section*{Acknowledgements}
We thank Abhimanyu Susobhanan for helpful discussions on pulsar timing and for insights into the inner workings of \texttt{PINT}. We are also grateful for the whole \texttt{PINT} and \texttt{Tempo2} community, as we have had many discussions with various people of those communities throughout the duration of this project.

\appendix

\section{Round-trip consistency}
\label{app:roundtrip}

Table~\ref{tab:roundtrip} gives the residual comparison of
Sec.~\ref{sec:roundtripconsistency} for every pulsar in the combination.
\texttt{PINT} and \texttt{Tempo2} each load the processed \texttt{.par} and
\texttt{.tim} with no fit, and the two residual series are compared on the
released TOAs. The tabulated RMS is the demeaned root-mean-square of their
difference, in nanoseconds. The median is $1.3\,\mathrm{ns}$ and the largest
value is $1.8\,\mathrm{ns}$, for PSR~J2017$+$0603.

\begin{longtable}{lrrrr}
\caption{Round-trip residual difference between \texttt{PINT} and
\texttt{Tempo2} for every pulsar in the combination. $N$ is the number of
PTAs and $N_{\rm TOA}$ the number of TOAs. The RMS and the maximum absolute
difference are in nanoseconds.}
\label{tab:roundtrip}\\
\hline
Pulsar & $N$ & $N_{\rm TOA}$ & RMS & max \\
\hline
\endfirsthead
\hline
Pulsar & $N$ & $N_{\rm TOA}$ & RMS & max \\
\hline
\endhead
\hline
\endfoot
B1855$+$09 & 3 & 4704 & 1.10 & 5.15 \\
B1937$+$21 & 3 & 9968 & 1.35 & 4.52 \\
B1953$+$29 & 2 & 1459 & 1.66 & 5.57 \\
J0023$+$0923 & 1 & 4373 & 1.51 & 3.05 \\
J0030$+$0451 & 2 & 3362 & 1.20 & 5.90 \\
J0034$-$0534 & 1 & 276 & 0.80 & 4.59 \\
J0218$+$4232 & 1 & 1196 & 1.09 & 7.26 \\
J0340$+$4130 & 1 & 3003 & 1.50 & 4.81 \\
J0437$-$4715 & 1 & 2455 & 1.12 & 4.66 \\
J0610$-$2100 & 1 & 1034 & 0.94 & 5.26 \\
J0613$-$0200 & 3 & 9200 & 1.27 & 10.49 \\
J0621$+$1002 & 1 & 673 & 1.11 & 8.39 \\
J0645$+$5158 & 1 & 2891 & 1.51 & 4.57 \\
J0711$-$6830 & 1 & 400 & 0.68 & 4.29 \\
J0751$+$1807 & 1 & 1491 & 1.07 & 9.01 \\
J0900$-$3144 & 1 & 875 & 1.13 & 5.70 \\
J0931$-$1902 & 1 & 712 & 1.23 & 3.96 \\
J1012$+$5307 & 2 & 13056 & 1.18 & 7.42 \\
J1022$+$1001 & 2 & 1270 & 0.89 & 4.26 \\
J1024$-$0719 & 3 & 5731 & 1.32 & 4.87 \\
J1045$-$4509 & 1 & 397 & 0.74 & 4.42 \\
J1455$-$3330 & 2 & 5507 & 1.29 & 4.82 \\
J1600$-$3053 & 3 & 8741 & 1.28 & 4.82 \\
J1603$-$7202 & 1 & 341 & 0.74 & 4.44 \\
J1614$-$2230 & 1 & 7323 & 1.40 & 4.96 \\
J1640$+$2224 & 2 & 3098 & 1.12 & 4.96 \\
J1643$-$1224 & 3 & 7998 & 1.29 & 4.84 \\
J1713$+$0747 & 3 & 16941 & 1.28 & 4.87 \\
J1721$-$2457 & 1 & 150 & 0.97 & 3.71 \\
J1730$-$2304 & 2 & 554 & 1.06 & 4.33 \\
J1738$+$0333 & 2 & 2885 & 1.36 & 4.57 \\
J1741$+$1351 & 1 & 1588 & 1.38 & 3.33 \\
J1744$-$1134 & 3 & 9637 & 1.20 & 4.68 \\
J1747$-$4036 & 1 & 2771 & 1.62 & 4.82 \\
J1751$-$2857 & 1 & 144 & 1.21 & 3.73 \\
J1801$-$1417 & 1 & 126 & 1.59 & 5.41 \\
J1802$-$2124 & 1 & 522 & 0.76 & 3.58 \\
J1804$-$2717 & 1 & 116 & 1.27 & 4.13 \\
J1832$-$0836 & 1 & 1131 & 1.51 & 3.91 \\
J1843$-$1113 & 1 & 224 & 1.23 & 5.24 \\
J1853$+$1303 & 2 & 1470 & 1.42 & 4.21 \\
J1903$+$0327 & 1 & 1802 & 1.51 & 4.64 \\
J1909$-$3744 & 3 & 11484 & 1.27 & 4.35 \\
J1910$+$1256 & 2 & 2743 & 1.31 & 3.50 \\
J1911$+$1347 & 1 & 140 & 1.29 & 4.24 \\
J1911$-$1114 & 1 & 130 & 1.48 & 5.02 \\
J1918$-$0642 & 2 & 9942 & 1.24 & 4.49 \\
J1923$+$2515 & 1 & 920 & 1.52 & 3.79 \\
J1944$+$0907 & 1 & 1696 & 1.44 & 4.23 \\
J1949$+$3106 & 1 & 1409 & 1.74 & 3.99 \\
J2010$-$1323 & 2 & 8057 & 1.31 & 4.55 \\
J2017$+$0603 & 1 & 1565 & 1.76 & 3.69 \\
J2033$+$1734 & 1 & 194 & 1.46 & 5.32 \\
J2043$+$1711 & 1 & 1382 & 1.58 & 4.51 \\
J2124$-$3358 & 2 & 893 & 0.97 & 4.48 \\
J2129$-$5721 & 1 & 270 & 0.75 & 4.21 \\
J2145$-$0750 & 3 & 8243 & 1.22 & 5.26 \\
J2214$+$3000 & 1 & 2514 & 1.55 & 2.84 \\
J2229$+$2643 & 1 & 316 & 1.30 & 4.95 \\
J2302$+$4442 & 1 & 3037 & 1.55 & 4.86 \\
J2317$+$1439 & 2 & 3175 & 1.11 & 4.88 \\
J2322$+$2057 & 1 & 229 & 1.35 & 4.95 \\

\end{longtable}

\section{Software corrections}
\label{app:softwarefixes}

The comparison in Sec.~\ref{sec:roundtripconsistency} measures the timing
model only where \texttt{PINT} and \texttt{Tempo2} implement the same model.
Several published files exposed places where they did not.

\paragraph{Solar-system ephemeris.}
\texttt{Tempo2} formed the Earth's Julian date in extended precision and
then passed it to the JPL reader as one \texttt{double}. Near the present
epoch that \texttt{double} has a step of about $40\,\mu\mathrm{s}$, and the
light-travel error along the Earth's velocity reaches $2\,\mathrm{ns}$ on
every TOA. The reader now receives the integer day and the fraction of the
day separately. The residuals in App.~\ref{app:roundtrip} use this form of
the lookup.

\paragraph{Binary delay.}
An ELL1 orbit may be stored as \texttt{PB} together with \texttt{FB2},
\texttt{FB3}, \ldots and no \texttt{FB1}. \texttt{Tempo2} treated a missing
\texttt{FB1} as the absence of the whole frequency series and dropped the
higher coefficients that the file did supply. A missing coefficient is zero.
The same files, and DDH solutions with a negative $h_3$, are accepted by the
companion change in \texttt{PINT}.
A Keplerian \texttt{BINARY T2} file was recognized by \texttt{PINT} as the
Blandford--Teukolsky model. With its default settings \texttt{Tempo2}
evaluates that file as Damour--Deruelle, and the recognition order now
selects DD.
For an ELL1H model, \texttt{Tempo2} evaluates the orthometric Shapiro delay
as Eq.~(28) of Freire and
Wex~\citep{freireOrthometricParametrizationShapiro2010}, in which the first
two harmonics are absorbed into the Roemer delay. \texttt{PINT} was
evaluating Eq.~(29), which keeps those harmonics in the Shapiro delay. The
two expressions differ by a gauge term. \texttt{PINT} now accepts either,
and \texttt{ell1h\_shapiro="absorbed"} selects the \texttt{Tempo2} choice.
The analytic derivative of the DDH delay with respect to $\varsigma$ was
large by a factor $1/T_\odot$. The delay itself was unaffected; the
design-matrix column, and with it the formal uncertainty on $\varsigma$, was.

\paragraph{Dispersion jumps.}
A positive \texttt{FDJUMPDM} decreased the dispersion delay in \texttt{PINT}
and increased it in \texttt{Tempo2}. The sign now agrees.
\texttt{libstempo} named the frequency jumps by the order of their lines in
the file, whereas \texttt{Tempo2} identifies each jump by its power of
frequency.

\paragraph{Names.}
The same quantity sometimes has two keywords. \texttt{Tempo2} writes the
ELL1H harmonic count as \texttt{NHARM}, and \texttt{PINT} writes
\texttt{NHARMS}. \texttt{PINT} reads only \texttt{NHARMS}. The defaults
differ as well: \texttt{PINT} floors the count at 7 when $h_4$ is present,
and \texttt{Tempo2} uses 4 when the keyword is absent. A shared file
therefore carries both lines, with one integer. A frequency jump is
\texttt{FDJUMP1} in a \texttt{Tempo2} file and \texttt{FD1JUMP} inside
\texttt{PINT}. \texttt{PINT} accepts the \texttt{Tempo2} spelling on input
and writes it back out. The dispersion-jump keyword is \texttt{FDJUMPDM} in
both packages; the disagreement on that parameter is the sign, above.

\paragraph{Evaluation order.}
\texttt{Tempo2} used to accumulate the frequency-dependent delay into the
dispersive delay, and the binary model was then evaluated at an epoch shifted
by that delay. \texttt{PINT} evaluates the binary first and applies the
frequency-dependent delay afterwards, as a correction for pulse-profile
evolution. \texttt{Tempo2} now adds the same delay as a phase, after the
binary.
On PSR~J1022$+$1001 in the NANOGrav 15-yr release, where the frequency-dependent
terms are unusually large, the old order left a residual difference of about
$480\,\mathrm{ns}$; removing the terms brought it to about $1\,\mathrm{ns}$.

\paragraph{Clocks and observatories.}
The Jodrell Bank Mark~II backends had no clock chain in \texttt{PINT}, which
then applied only a constant of about $27\,\mu\mathrm{s}$. On the EPTA
observations of PSR~J1713$+$0747 the residual difference fell from about
$660\,\mathrm{ns}$ to a few nanoseconds once those backends followed the
Lovell chains.
The telescope coordinates disagreed as well, for two different reasons.
In June 2026 \texttt{PINT} moved the GMRT to the InPTA position, a shift of
$750\,\mathrm{m}$, about $2.5\,\mu\mathrm{s}$ of light travel.
Effelsberg was already on its VLBI position in \texttt{PINT}; the coordinate
shipped with \texttt{Tempo2} was an older value, $2.8\,\mathrm{m}$ away,
about $9\,\mathrm{ns}$. \texttt{Tempo2} now uses both of the \texttt{PINT}
positions.
\texttt{libstempo} exposed the Earth-center velocity in the array slots that
\texttt{Tempo2} leaves at zero, and did not expose the velocity of the
telescope. The omitted rotation of the Earth is about $100\,\mathrm{ns}$ of
dispersion delay once the delay is referred to the observatory.

\paragraph{Reference frames.}
Transporting an ecliptic astrometric uncertainty to equatorial coordinates
treated $(\sigma_\lambda, \sigma_\beta)$ as a signed vector. The marginal
uncertainties after a rotation on the sky are the diagonal of the rotated
covariance, and the vector treatment could return a negative uncertainty.
In \texttt{Vela} the DDK annual-parallax term built its basis from a pulsar
direction in the ICRS and a node angle in the ecliptic. Both vectors are now
taken in the frame of the node.

\paragraph{Timescale conversion.}
\texttt{PINT} converted \texttt{TCB} to \texttt{TDB} from the legacy IFTE
common origin. The revised conversion holds a supported deterministic model
to about a nanosecond, with no refit of the published solution.

\section{Binary models}
\label{app:binarymodels}

\begin{table}[tb]
\caption{Binary-model parameters. A blank \texttt{.par} name is a quantity
used in the definitions; an alias is given in the meaning. DD$^*$
denotes DD, DDH, and DDK; ELL1$^*$ denotes ELL1 and ELL1H. $T_\odot =
GM_\odot/c^3$.}
\label{tab:binaryparams}
\begin{tabular}{ll>{\raggedright\arraybackslash}p{0.40\columnwidth}l}
\hline
\texttt{.par} & Symbol & Meaning & Models \\
\hline
\texttt{PB} & $P_b$ & orbital period & all \\
& $a_p$ & semi-major axis of the pulsar orbit & \\
& $a_p\sin i$ & projected semi-major axis & \\
\texttt{A1} & $x$ & $x = a_p\sin i/c$ & all \\
\texttt{ECC} & $e$ & eccentricity; \texttt{E} is an alias & BT, DD$^*$ \\
\texttt{OM} & $\omega$ & longitude of periastron & BT, DD$^*$ \\
\texttt{T0} & $T_0$ & epoch of periastron passage & BT, DD$^*$ \\
\texttt{EPS1} & $\varepsilon_1$ & $e\sin\omega$ & ELL1$^*$ \\
\texttt{EPS2} & $\varepsilon_2$ & $e\cos\omega$ & ELL1$^*$ \\
\texttt{TASC} & $T_{\rm asc}$ & epoch of ascending-node passage & ELL1$^*$ \\
\texttt{PBDOT} & $\dot P_b$ & orbital-period derivative & all \\
\texttt{XDOT} & $\dot x$ & derivative of $x$; \texttt{A1DOT} is an alias & all \\
\texttt{EPS1DOT} & $\dot\varepsilon_1$ & derivative of $\varepsilon_1$ & ELL1$^*$ \\
\texttt{EPS2DOT} & $\dot\varepsilon_2$ & derivative of $\varepsilon_2$ & ELL1$^*$ \\
\texttt{OMDOT} & $\dot\omega$ & periastron advance & BT, DD$^*$ \\
\texttt{GAMMA} & $\gamma$ & Einstein-delay amplitude & BT, DD$^*$ \\
\texttt{M2} & $M_2$ & companion mass & DD, DDK, ELL1 \\
\texttt{SINI} & $\sin i$ & sine of the orbital inclination; DDK derives it from \texttt{KIN} & DD, ELL1 \\
\texttt{H3} & $h_3$ & orthometric Shapiro amplitude, $T_\odot M_2\varsigma^3$ & DDH, ELL1H \\
\texttt{H4} & $h_4$ & fourth orthometric harmonic, $T_\odot M_2\varsigma^4$ & ELL1H \\
\texttt{STIG} & $\varsigma$ & orthometric ratio, $\sin i/(1+|\cos i|)$ & DDH, ELL1H \\
\texttt{KIN} & $i$ & orbital inclination & DDK \\
\texttt{KOM} & $\Omega$ & longitude of the ascending node & DDK \\
\hline
\end{tabular}
\end{table}

This appendix describes what Sec.~\ref{sec:binarymodels} leaves out: the
binary models found in published PTA data, the reasons the PTAs differ in
their choice, and the steps by which \texttt{MetaPulsar} brings the models of
one pulsar into a single form that \texttt{PINT} and \texttt{Tempo2} evaluate
identically.

\subsection{The models}

The binary models that show up in published PTA data all come from the same
line of development, and Table~\ref{tab:binaryparams} lists the parameters
that go with them. \citet{blandfordArrivaltimeAnalysisPulsar1976} wrote the
orbit down as a Keplerian ellipse, and put in the Einstein delay and the
secular drifts of the orbital elements phenomenologically. This is the BT
model. \citet{damourGeneralRelativisticCelestial1985,damourGeneralRelativisticCelestial1986}
then derived the timing formula properly from post-Newtonian celestial
mechanics, which adds the Shapiro delay, the aberration delay, and the
relativistic deformation of the orbit. The DD model that came out of that
work is what most binary models in use today are built on. Its Keplerian
elements are $(P_b, x, e, \omega, T_0)$, and the relativistic effects come
in as post-Keplerian parameters. What a PTA dataset actually measures of
these are the secular drifts $(\dot P_b, \dot\omega, \dot x)$ and the
Shapiro delay, parametrized by the companion mass $M_2$ and $\sin i$.

Almost every binary in a PTA is a millisecond pulsar with a white-dwarf
companion, with an eccentricity somewhere between $10^{-7}$ and $10^{-4}$.
DD does not handle such orbits well. As $e$ goes to zero the periastron
stops being defined, $\omega$ and $T_0$ become degenerate, and the design
matrix of the DD model becomes singular.
\citet{langePrecisionTimingMeasurements2001} dealt with this by introducing
the Laplace--Lagrange parameters $\varepsilon_1 = e\sin\omega$ and
$\varepsilon_2 = e\cos\omega$, together with the epoch of ascending node
$T_{\rm asc} = T_0 - \omega P_b/2\pi$. All three of these stay well defined
at $e=0$. This is an exact reparametrization of the same orbit, so it is
tempting to convert $(\varepsilon_1,\varepsilon_2,T_{\rm asc})$ back to
$(e,\omega,T_0)$ and to evaluate the DD delay as usual. However, this does not work in practice. The inverse transformation has a coordinate singularity at $e=0$, which
means the conversion is numerically ill-defined in exactly the orbits where
we would want to use it. ELL1 is therefore a delay model of its own: the
R{\o}mer delay is expanded in powers of $e$ and written directly in terms of
$\varepsilon_1$, $\varepsilon_2$, and the orbital phase measured from
$T_{\rm asc}$. Because that expansion is cut off, ELL1 is an approximation,
and it is only a good one when the terms that were left out lie below the
timing precision.

Later the Shapiro delay was given its own parametrization in both of these
models~\citep{freireOrthometricParametrizationShapiro2010}, which is where
DDH and ELL1H come from. The orthometric ratio
$\varsigma = \sin i/(1+|\cos i|)$ and the amplitude
$h_3 = T_\odot M_2\,\varsigma^3$ are far less correlated than $M_2$ and
$\sin i$, and $h_3$ can still be measured in cases where the Shapiro delay
is too weak to constrain $M_2$ and $\sin i$ separately. For nearby pulsars
in wide orbits there is one more effect that becomes measurable: the
annual-orbital parallax, and the apparent drifts in $x$ and $\omega$ that
proper motion
induces~\citep{kopeikinPossibleImplicationsOrbital1995,kopeikinProperMotionBinary1996}.
The DDK model includes these, through the inclination $i$ and the longitude
of the ascending node $\Omega$.

On top of all of the above, \texttt{Tempo2} ships a general T2
model~\citep{edwardsTempo2NewPulsar2006}. The T2 model is different from the models above. It is a single routine that holds the delay formulae of the models above and
chooses between them from the parameters that are present in the
\texttt{.par} file. Given an eccentricity it follows the DD formulae, and it
adds the Kopeikin terms when $i$ and $\Omega$ are present. Given
$\varepsilon_1$ and $\varepsilon_2$ instead it follows the ELL1 expansion, and
it then takes the Shapiro delay from $(M_2, \sin i)$ or from one of several
orthometric forms, again depending on which parameters appear. A
\texttt{.par} file that declares T2 therefore does not name its delay model:
the model is decided, parameter by parameter, when the file is read.

\subsection{Why the PTAs differ}

Which model ends up in a \texttt{.par} file is decided by the data, and that
is why two PTAs routinely publish different binary models for the same
pulsar. When the error from truncating the ELL1 expansion lies below the
timing precision, ELL1 is the model to use, because $\omega$ and $T_0$ are
then poorly constrained and strongly correlated. When the eccentricity
itself can be measured, DD is used instead. The orthometric parameters are
used when the Shapiro delay is detected only weakly, and $(M_2, \sin i)$
when it is detected strongly. DDK is used when the Kopeikin terms are large
enough to matter (currently only PSR~J1713$+$0747). So a PTA with a longer baseline, or with better timing
precision, sometimes uses a different model than another PTA for one and the same
pulsar.

\subsection{What \texttt{MetaPulsar} does}

\texttt{MetaPulsar} brings the binary models of one pulsar into a single,
explicitly named model in three steps, taken in this order. $42$ of the $62$
pulsars of the combination in this paper are binaries.

\emph{Resolving T2.} Every T2 declaration is replaced by the name of the
concrete model that its parameters select, such as ELL1, ELL1H, DD, or DDK.
\texttt{PINT} has no T2 model and makes this choice itself when it reads such
a file; writing the result into the file means that both packages are handed
the same named model. One parameter convention changes when the binary model is converted from T2: T2 measures the
Kopeikin angles in the IAU convention, and the DDK model of both packages
uses that of \citet{damourStrongFieldTests1992}, so a T2 model
that resolves to DDK has its angles mapped, $i \rightarrow 180^\circ - i$ and
$\Omega \rightarrow 90^\circ - \Omega$.

\emph{Adopting the reference model.} The binary model of the reference PTA,
with its parameter values, replaces the binary model of every other dataset
of the pulsar. No map between the parametrizations of two PTAs is needed for
this. The values serve as the expansion point of the shared model
(Sec.~\ref{sec:timingmodelvalues}), and the data of the other PTAs then
constrain the parameters of the reference. A dataset whose own release did
not model the Shapiro delay, for example, acquires the Shapiro parameters of
the reference.

\emph{Converting ELL1(H) where the packages disagree.} One difference is between the timing packages, and not between the models. \texttt{Tempo2}
follows \citet{langePrecisionTimingMeasurements2001} and expands ELL1(H) to
first order in $e$, so the terms it neglects are of order $x e^2$.
\texttt{PINT} keeps terms up to third
order~\citep{zhuTestsGravitationalSymmetries2019,luoPINTModernSoftware2021},
so it only neglects terms of order $x e^4$. To see when this matters, take
$x = 10\,\mathrm{lt\text{-}s}$: then $x e^2$ reaches $1\,\mathrm{ns}$ already
at $e \approx 10^{-5}$. For a pulsar at that eccentricity, ELL1(H) is a
perfectly adequate model in \texttt{PINT} and not in \texttt{Tempo2}, and
the same \texttt{.par} file gives different delays in the two packages, in
violation of requirement (a) of Sec.~\ref{sec:fullspecification}.
\texttt{MetaPulsar} therefore tests every ELL1 and ELL1H model whenever the
combined model has to be evaluated by both packages, which is the setting of
every combination in this paper. The test is a conservative estimate of the
neglected terms, $x e^2 + n_b x^2 e/2$ with $n_b = 2\pi/P_b$, taken at the
largest $x$ and $e$ that the model reaches over the data span. Where it
exceeds $1\,\mathrm{ns}$ the shared model is converted to
DD~\citep{damourGeneralRelativisticCelestial1986}, or to DDH when the source
is orthometric. In this paper that is the case for five pulsars,
PSRs~J0900$-$3144, J1741$+$1351, J1802$-$2124, J1918$-$0642, and
J1949$+$3106, all of which go to DD.

The conversion maps $(\varepsilon_1, \varepsilon_2, T_{\rm asc})$ to
$(e, \omega, T_0)$, with the correction terms that account for what each
delay formula absorbs into $x$ and into the epoch. This may seem to contradict the above, since ELL1 was introduced to avoid this transformation. However, the map is singular only at $e=0$, and the test fires only where the eccentricity is
large enough to matter: in the five pulsars above it is detected at between
$17$ and $1100$ standard deviations, so the periastron is well defined. What
the conversion does bring back is the correlation between $\omega$ and $T_0$,
which grows as $e$ shrinks. That is harmless for an analysis that
marginalizes over the timing model, but it means that only their combination
is well measured: in Fig.~\ref{fig:j0613_triple} the shift of $T_0$ for
PSR~J1600$-$3053, which NANOGrav releases as DD, is accompanied by an
identical shift of $\omega$.

Every conversion is verified by comparing the predicted binary delays of the
source model and of the converted model, and a conversion that fails the
check is refused. An ELL1H source that carries $h_3$ and neither $\varsigma$
nor $h_4$ is refused for a different reason. Both packages evaluate that
file as the single harmonic $-\frac{4}{3}h_3\sin 3\Phi$. DDH has no such
truncation: it requires $h_3$ and $\varsigma$ and evaluates the exact
Shapiro delay, and the terms absorbed into $x$ and the epoch cannot be
removed without $\varsigma$. No pulsar in this paper is in that state.

\bibliography{sample}

\begin{thebibliography}{48}%
\makeatletter
\providecommand \@ifxundefined [1]{%
 \@ifx{#1\undefined}
}%
\providecommand \@ifnum [1]{%
 \ifnum #1\expandafter \@firstoftwo
 \else \expandafter \@secondoftwo
 \fi
}%
\providecommand \@ifx [1]{%
 \ifx #1\expandafter \@firstoftwo
 \else \expandafter \@secondoftwo
 \fi
}%
\providecommand \natexlab [1]{#1}%
\providecommand \enquote  [1]{``#1''}%
\providecommand \bibnamefont  [1]{#1}%
\providecommand \bibfnamefont [1]{#1}%
\providecommand \citenamefont [1]{#1}%
\providecommand \href@noop [0]{\@secondoftwo}%
\providecommand \href [0]{\begingroup \@sanitize@url \@href}%
\providecommand \@href[1]{\@@startlink{#1}\@@href}%
\providecommand \@@href[1]{\endgroup#1\@@endlink}%
\providecommand \@sanitize@url [0]{\catcode `\\12\catcode `\$12\catcode
  `\&12\catcode `\#12\catcode `\^12\catcode `\_12\catcode `\%12\relax}%
\providecommand \@@startlink[1]{}%
\providecommand \@@endlink[0]{}%
\providecommand \url  [0]{\begingroup\@sanitize@url \@url }%
\providecommand \@url [1]{\endgroup\@href {#1}{\urlprefix }}%
\providecommand \urlprefix  [0]{URL }%
\providecommand \Eprint [0]{\href }%
\providecommand \doibase [0]{https://doi.org/}%
\providecommand \selectlanguage [0]{\@gobble}%
\providecommand \bibinfo  [0]{\@secondoftwo}%
\providecommand \bibfield  [0]{\@secondoftwo}%
\providecommand \translation [1]{[#1]}%
\providecommand \BibitemOpen [0]{}%
\providecommand \bibitemStop [0]{}%
\providecommand \bibitemNoStop [0]{.\EOS\space}%
\providecommand \EOS [0]{\spacefactor3000\relax}%
\providecommand \BibitemShut  [1]{\csname bibitem#1\endcsname}%
\let\auto@bib@innerbib\@empty
\bibitem [{\citenamefont {Hellings}\ and\ \citenamefont
  {Downs}(1983)}]{HellingsDowns1983}%
  \BibitemOpen
  \bibfield  {author} {\bibinfo {author} {\bibfnamefont {R.}~\bibnamefont
  {Hellings}}\ and\ \bibinfo {author} {\bibfnamefont {G.}~\bibnamefont
  {Downs}},\ }\bibfield  {title} {\bibinfo {title} {{\color{Gray}Upper limits
  on the isotropic gravitational radiation background from pulsar timing
  analysis}},\ }\href {https://doi.org/10.1086/183954} {\bibfield  {journal}
  {\bibinfo  {journal} {The Astrophysical Journal}\ }\textbf {\bibinfo {volume}
  {265}},\ \bibinfo {pages} {L39} (\bibinfo {year} {1983})}\BibitemShut
  {NoStop}%
\bibitem [{\citenamefont {Abbott}\ \emph {et~al.}(2016)\citenamefont {Abbott},
  \citenamefont {Abbott}, \citenamefont {Abbott}, \citenamefont {Abernathy},
  \citenamefont {Acernese}, \citenamefont {Ackley}, \citenamefont {Adams},
  \citenamefont {Adams}, \citenamefont {Addesso}, \citenamefont {Adhikari}
  \emph {et~al.}}]{abbott2016observation}%
  \BibitemOpen
  \bibfield  {author} {\bibinfo {author} {\bibfnamefont {B.~P.}\ \bibnamefont
  {Abbott}}, \bibinfo {author} {\bibfnamefont {R.}~\bibnamefont {Abbott}},
  \bibinfo {author} {\bibfnamefont {T.~D.}\ \bibnamefont {Abbott}}, \emph
  {et~al.},\ }\bibfield  {title} {\bibinfo {title} {{\color{Gray}{Observation}
  of {Gravitational} {Waves} from a {Binary} {Black} {Hole} {Merger}}},\ }\href
  {https://doi.org/10.1103/PhysRevLett.116.061102} {\bibfield  {journal}
  {\bibinfo  {journal} {Physical Review Letters}\ }\textbf {\bibinfo {volume}
  {116}},\ \bibinfo {pages} {061102} (\bibinfo {year} {2016})},\ \Eprint
  {https://arxiv.org/abs/1602.03837} {arXiv:1602.03837 [gr-qc]} \BibitemShut
  {NoStop}%
\bibitem [{\citenamefont {Rajagopal}\ and\ \citenamefont
  {Romani}(1995)}]{rajagopal1994ultra}%
  \BibitemOpen
  \bibfield  {author} {\bibinfo {author} {\bibfnamefont {M.}~\bibnamefont
  {Rajagopal}}\ and\ \bibinfo {author} {\bibfnamefont {R.~W.}\ \bibnamefont
  {Romani}},\ }\bibfield  {title} {\bibinfo {title} {{\color{Gray}{Ultra}-{Low}
  {Frequency} {Gravitational} {Radiation} from {Massive} {Black} {Hole}
  {Binaries}}},\ }\href {https://doi.org/10.1086/175813} {\bibfield  {journal}
  {\bibinfo  {journal} {The Astrophysical Journal}\ }\textbf {\bibinfo {volume}
  {446}},\ \bibinfo {pages} {543} (\bibinfo {year} {1995})},\ \Eprint
  {https://arxiv.org/abs/astro-ph/9412038} {arXiv:astro-ph/9412038}
  \BibitemShut {NoStop}%
\bibitem [{\citenamefont {Jaffe}\ and\ \citenamefont
  {Backer}(2003)}]{jaffe2003gravitational}%
  \BibitemOpen
  \bibfield  {author} {\bibinfo {author} {\bibfnamefont {A.~H.}\ \bibnamefont
  {Jaffe}}\ and\ \bibinfo {author} {\bibfnamefont {D.~C.}\ \bibnamefont
  {Backer}},\ }\bibfield  {title} {\bibinfo {title} {{\color{Gray}Gravitational
  waves probe the coalescence rate of massive black hole binaries}},\ }\href
  {https://doi.org/10.1086/345443} {\bibfield  {journal} {\bibinfo  {journal}
  {The Astrophysical Journal}\ }\textbf {\bibinfo {volume} {583}},\ \bibinfo
  {pages} {616} (\bibinfo {year} {2003})},\ \Eprint
  {https://arxiv.org/abs/astro-ph/0210148} {arXiv:astro-ph/0210148}
  \BibitemShut {NoStop}%
\bibitem [{\citenamefont {Wyithe}\ and\ \citenamefont
  {Loeb}(2003)}]{wyithe2003low}%
  \BibitemOpen
  \bibfield  {author} {\bibinfo {author} {\bibfnamefont {J.~S.~B.}\
  \bibnamefont {Wyithe}}\ and\ \bibinfo {author} {\bibfnamefont
  {A.}~\bibnamefont {Loeb}},\ }\bibfield  {title} {\bibinfo {title}
  {{\color{Gray}{Low-Frequency} {Gravitational} {Waves} from {Massive} {Black}
  {Hole} {Binaries}: {Predictions} for {LISA} and {Pulsar} {Timing}
  {Arrays}}},\ }\href {https://doi.org/10.1086/375187} {\bibfield  {journal}
  {\bibinfo  {journal} {The Astrophysical Journal}\ }\textbf {\bibinfo {volume}
  {590}},\ \bibinfo {pages} {691} (\bibinfo {year} {2003})},\ \Eprint
  {https://arxiv.org/abs/astro-ph/0211556} {arXiv:astro-ph/0211556}
  \BibitemShut {NoStop}%
\bibitem [{\citenamefont {Sesana}\ \emph {et~al.}(2008)\citenamefont {Sesana},
  \citenamefont {Vecchio},\ and\ \citenamefont
  {Colacino}}]{sesana2008stochastic}%
  \BibitemOpen
  \bibfield  {author} {\bibinfo {author} {\bibfnamefont {A.}~\bibnamefont
  {Sesana}}, \bibinfo {author} {\bibfnamefont {A.}~\bibnamefont {Vecchio}},\
  and\ \bibinfo {author} {\bibfnamefont {C.~N.}\ \bibnamefont {Colacino}},\
  }\bibfield  {title} {\bibinfo {title} {{\color{Gray}The stochastic
  gravitational-wave background from massive black hole binary systems:
  implications for observations with Pulsar Timing Arrays}},\ }\href
  {https://doi.org/10.1111/j.1365-2966.2008.13682.x} {\bibfield  {journal}
  {\bibinfo  {journal} {Monthly Notices of the Royal Astronomical Society}\
  }\textbf {\bibinfo {volume} {390}},\ \bibinfo {pages} {192} (\bibinfo {year}
  {2008})},\ \Eprint {https://arxiv.org/abs/0804.4476} {arXiv:0804.4476
  [astro-ph]} \BibitemShut {NoStop}%
\bibitem [{\citenamefont {Rosado}\ \emph {et~al.}(2015)\citenamefont {Rosado},
  \citenamefont {Sesana},\ and\ \citenamefont {Gair}}]{rosado2015expected}%
  \BibitemOpen
  \bibfield  {author} {\bibinfo {author} {\bibfnamefont {P.~A.}\ \bibnamefont
  {Rosado}}, \bibinfo {author} {\bibfnamefont {A.}~\bibnamefont {Sesana}},\
  and\ \bibinfo {author} {\bibfnamefont {J.}~\bibnamefont {Gair}},\ }\bibfield
  {title} {\bibinfo {title} {{\color{Gray}Expected properties of the first
  gravitational wave signal detected with pulsar timing arrays}},\ }\href
  {https://doi.org/10.1093/mnras/stv1098} {\bibfield  {journal} {\bibinfo
  {journal} {Monthly Notices of the Royal Astronomical Society}\ }\textbf
  {\bibinfo {volume} {451}},\ \bibinfo {pages} {2417} (\bibinfo {year}
  {2015})},\ \Eprint {https://arxiv.org/abs/1503.04803} {arXiv:1503.04803
  [astro-ph.HE]} \BibitemShut {NoStop}%
\bibitem [{\citenamefont {Antoniadis}\ \emph {et~al.}(2023)\citenamefont
  {Antoniadis}, \citenamefont {Arumugam}, \citenamefont {Arumugam},
  \citenamefont {Babak}, \citenamefont {Bagchi}, \citenamefont {Nielsen},
  \citenamefont {Bassa}, \citenamefont {Bathula}, \citenamefont {Berthereau},
  \citenamefont {Bonetti} \emph {et~al.}}]{antoniadis2023second}%
  \BibitemOpen
  \bibfield  {author} {\bibinfo {author} {\bibfnamefont {J.}~\bibnamefont
  {Antoniadis}}, \bibinfo {author} {\bibfnamefont {P.}~\bibnamefont
  {Arumugam}}, \bibinfo {author} {\bibfnamefont {S.}~\bibnamefont {Arumugam}},
  \emph {et~al.},\ }\bibfield  {title} {\bibinfo {title} {{\color{Gray}The
  second data release from the European Pulsar Timing Array-III. Search for
  gravitational wave signals}},\ }\href
  {https://doi.org/10.1051/0004-6361/202346844} {\bibfield  {journal} {\bibinfo
   {journal} {Astronomy \& Astrophysics}\ }\textbf {\bibinfo {volume} {678}},\
  \bibinfo {pages} {A50} (\bibinfo {year} {2023})},\ \Eprint
  {https://arxiv.org/abs/2306.16214} {arXiv:2306.16214 [astro-ph.HE]}
  \BibitemShut {NoStop}%
\bibitem [{\citenamefont {Agazie}\ \emph
  {et~al.}(2023{\natexlab{a}})\citenamefont {Agazie}, \citenamefont
  {Anumarlapudi}, \citenamefont {Archibald}, \citenamefont {Arzoumanian},
  \citenamefont {Baker}, \citenamefont {B{\'e}csy}, \citenamefont {Blecha},
  \citenamefont {Brazier}, \citenamefont {Brook}, \citenamefont {Burke-Spolaor}
  \emph {et~al.}}]{agazie2023nanograv}%
  \BibitemOpen
  \bibfield  {author} {\bibinfo {author} {\bibfnamefont {G.}~\bibnamefont
  {Agazie}}, \bibinfo {author} {\bibfnamefont {A.}~\bibnamefont
  {Anumarlapudi}}, \bibinfo {author} {\bibfnamefont {A.~M.}\ \bibnamefont
  {Archibald}}, \emph {et~al.},\ }\bibfield  {title} {\bibinfo {title}
  {{\color{Gray}The {NANOGrav} 15 yr {Data} {Set}: {Evidence} for a
  {Gravitational}-wave {Background}}},\ }\href
  {https://doi.org/10.3847/2041-8213/acdac6} {\bibfield  {journal} {\bibinfo
  {journal} {The Astrophysical Journal Letters}\ }\textbf {\bibinfo {volume}
  {951}},\ \bibinfo {pages} {L8} (\bibinfo {year} {2023}{\natexlab{a}})},\
  \Eprint {https://arxiv.org/abs/2306.16213} {arXiv:2306.16213 [astro-ph.HE]}
  \BibitemShut {NoStop}%
\bibitem [{\citenamefont {Reardon}\ \emph {et~al.}(2023)\citenamefont
  {Reardon}, \citenamefont {Zic}, \citenamefont {Shannon}, \citenamefont
  {Hobbs}, \citenamefont {Bailes}, \citenamefont {Di~Marco}, \citenamefont
  {Kapur}, \citenamefont {Rogers}, \citenamefont {Thrane}, \citenamefont
  {Askew} \emph {et~al.}}]{reardon2023search}%
  \BibitemOpen
  \bibfield  {author} {\bibinfo {author} {\bibfnamefont {D.~J.}\ \bibnamefont
  {Reardon}}, \bibinfo {author} {\bibfnamefont {A.}~\bibnamefont {Zic}},
  \bibinfo {author} {\bibfnamefont {R.~M.}\ \bibnamefont {Shannon}}, \emph
  {et~al.},\ }\bibfield  {title} {\bibinfo {title} {{\color{Gray}Search for an
  isotropic gravitational-wave background with the Parkes Pulsar Timing
  Array}},\ }\href {https://doi.org/10.3847/2041-8213/acdd02} {\bibfield
  {journal} {\bibinfo  {journal} {The Astrophysical Journal Letters}\ }\textbf
  {\bibinfo {volume} {951}},\ \bibinfo {pages} {L6} (\bibinfo {year} {2023})},\
  \Eprint {https://arxiv.org/abs/2306.16215} {arXiv:2306.16215 [astro-ph.HE]}
  \BibitemShut {NoStop}%
\bibitem [{\citenamefont {Xu}\ \emph {et~al.}(2023)\citenamefont {Xu},
  \citenamefont {Chen}, \citenamefont {Guo}, \citenamefont {Jiang},
  \citenamefont {Wang}, \citenamefont {Xu}, \citenamefont {Xue}, \citenamefont
  {Caballero}, \citenamefont {Yuan}, \citenamefont {Xu} \emph
  {et~al.}}]{xu2023searching}%
  \BibitemOpen
  \bibfield  {author} {\bibinfo {author} {\bibfnamefont {H.}~\bibnamefont
  {Xu}}, \bibinfo {author} {\bibfnamefont {S.}~\bibnamefont {Chen}}, \bibinfo
  {author} {\bibfnamefont {Y.}~\bibnamefont {Guo}}, \emph {et~al.},\ }\bibfield
   {title} {\bibinfo {title} {{\color{Gray}Searching for the {Nano}-{Hertz}
  {Stochastic} {Gravitational} {Wave} {Background} with the {Chinese} {Pulsar}
  {Timing} {Array} {Data} {Release} I}},\ }\href
  {https://doi.org/10.1088/1674-4527/acdfa5} {\bibfield  {journal} {\bibinfo
  {journal} {Research in Astronomy and Astrophysics}\ }\textbf {\bibinfo
  {volume} {23}},\ \bibinfo {pages} {075024} (\bibinfo {year} {2023})},\
  \Eprint {https://arxiv.org/abs/2306.16216} {arXiv:2306.16216 [astro-ph.HE]}
  \BibitemShut {NoStop}%
\bibitem [{\citenamefont {Miles}\ \emph {et~al.}(2025)\citenamefont {Miles},
  \citenamefont {Shannon}, \citenamefont {Reardon}, \citenamefont {Bailes},
  \citenamefont {Champion}, \citenamefont {Geyer}, \citenamefont {Gitika},
  \citenamefont {Grunthal}, \citenamefont {Keith}, \citenamefont {Kramer} \emph
  {et~al.}}]{miles2025meerkat}%
  \BibitemOpen
  \bibfield  {author} {\bibinfo {author} {\bibfnamefont {M.~T.}\ \bibnamefont
  {Miles}}, \bibinfo {author} {\bibfnamefont {R.~M.}\ \bibnamefont {Shannon}},
  \bibinfo {author} {\bibfnamefont {D.~J.}\ \bibnamefont {Reardon}}, \emph
  {et~al.},\ }\bibfield  {title} {\bibinfo {title} {{\color{Gray}The MeerKAT
  Pulsar Timing Array: The first search for gravitational waves with the
  MeerKAT radio telescope}},\ }\href {https://doi.org/10.1093/mnras/stae2571}
  {\bibfield  {journal} {\bibinfo  {journal} {Monthly Notices of the Royal
  Astronomical Society}\ }\textbf {\bibinfo {volume} {536}},\ \bibinfo {pages}
  {1489} (\bibinfo {year} {2025})},\ \Eprint {https://arxiv.org/abs/2412.01153}
  {arXiv:2412.01153 [astro-ph.HE]} \BibitemShut {NoStop}%
\bibitem [{\citenamefont {Siemens}\ \emph {et~al.}(2013)\citenamefont
  {Siemens}, \citenamefont {Ellis}, \citenamefont {Jenet},\ and\ \citenamefont
  {Romano}}]{siemens2013stochastic}%
  \BibitemOpen
  \bibfield  {author} {\bibinfo {author} {\bibfnamefont {X.}~\bibnamefont
  {Siemens}}, \bibinfo {author} {\bibfnamefont {J.}~\bibnamefont {Ellis}},
  \bibinfo {author} {\bibfnamefont {F.}~\bibnamefont {Jenet}},\ and\ \bibinfo
  {author} {\bibfnamefont {J.~D.}\ \bibnamefont {Romano}},\ }\bibfield  {title}
  {\bibinfo {title} {{\color{Gray}The stochastic background: scaling laws and
  time to detection for pulsar timing arrays}},\ }\href
  {https://doi.org/10.1088/0264-9381/30/22/224015} {\bibfield  {journal}
  {\bibinfo  {journal} {Classical and Quantum Gravity}\ }\textbf {\bibinfo
  {volume} {30}},\ \bibinfo {pages} {224015} (\bibinfo {year} {2013})},\
  \Eprint {https://arxiv.org/abs/1305.3196} {arXiv:1305.3196 [astro-ph.IM]}
  \BibitemShut {NoStop}%
\bibitem [{\citenamefont {Perera}\ \emph {et~al.}(2019)\citenamefont {Perera},
  \citenamefont {DeCesar}, \citenamefont {Demorest}, \citenamefont {Kerr},
  \citenamefont {Lentati}, \citenamefont {Nice}, \citenamefont {Os{\l}owski},
  \citenamefont {Ransom}, \citenamefont {Keith}, \citenamefont {Arzoumanian}
  \emph {et~al.}}]{perera2019international}%
  \BibitemOpen
  \bibfield  {author} {\bibinfo {author} {\bibfnamefont {B.}~\bibnamefont
  {Perera}}, \bibinfo {author} {\bibfnamefont {M.}~\bibnamefont {DeCesar}},
  \bibinfo {author} {\bibfnamefont {P.}~\bibnamefont {Demorest}}, \emph
  {et~al.},\ }\bibfield  {title} {\bibinfo {title} {{\color{Gray}The
  {International} {Pulsar} {Timing} {Array}: second data release}},\ }\href
  {https://doi.org/10.1093/mnras/stz2857} {\bibfield  {journal} {\bibinfo
  {journal} {Monthly Notices of the Royal Astronomical Society}\ }\textbf
  {\bibinfo {volume} {490}},\ \bibinfo {pages} {4666} (\bibinfo {year}
  {2019})},\ \Eprint {https://arxiv.org/abs/1909.04534} {arXiv:1909.04534
  [astro-ph.HE]} \BibitemShut {NoStop}%
\bibitem [{\citenamefont {Antoniadis}\ \emph {et~al.}(2022)\citenamefont
  {Antoniadis}, \citenamefont {Arzoumanian}, \citenamefont {Babak},
  \citenamefont {Bailes}, \citenamefont {Bak~Nielsen}, \citenamefont {Baker},
  \citenamefont {Bassa}, \citenamefont {B{\'e}csy}, \citenamefont {Berthereau},
  \citenamefont {Bonetti} \emph {et~al.}}]{antoniadis2022international}%
  \BibitemOpen
  \bibfield  {author} {\bibinfo {author} {\bibfnamefont {J.}~\bibnamefont
  {Antoniadis}}, \bibinfo {author} {\bibfnamefont {Z.}~\bibnamefont
  {Arzoumanian}}, \bibinfo {author} {\bibfnamefont {S.}~\bibnamefont {Babak}},
  \emph {et~al.},\ }\bibfield  {title} {\bibinfo {title} {{\color{Gray}The
  International Pulsar Timing Array second data release: Search for an
  isotropic gravitational wave background}},\ }\href
  {https://doi.org/10.1093/mnras/stab3418} {\bibfield  {journal} {\bibinfo
  {journal} {Monthly Notices of the Royal Astronomical Society}\ }\textbf
  {\bibinfo {volume} {510}},\ \bibinfo {pages} {4873} (\bibinfo {year}
  {2022})},\ \Eprint {https://arxiv.org/abs/2201.03980} {arXiv:2201.03980
  [astro-ph.HE]} \BibitemShut {NoStop}%
\bibitem [{\citenamefont {Arzoumanian}\ \emph {et~al.}(2018)\citenamefont
  {Arzoumanian}, \citenamefont {Brazier}, \citenamefont {{Burke-Spolaor}},
  \citenamefont {Chamberlin}, \citenamefont {Chatterjee}, \citenamefont
  {Christy}, \citenamefont {Cordes}, \citenamefont {Cornish}, \citenamefont
  {Crawford}, \citenamefont {Cromartie}, \citenamefont {Crowter}, \citenamefont
  {DeCesar}, \citenamefont {Demorest}, \citenamefont {Dolch}, \citenamefont
  {Ellis}, \citenamefont {Ferdman}, \citenamefont {Ferrara}, \citenamefont
  {Fonseca}, \citenamefont {{Garver-Daniels}}, \citenamefont {Gentile},
  \citenamefont {Halmrast}, \citenamefont {Huerta}, \citenamefont {Jenet},
  \citenamefont {Jessup}, \citenamefont {Jones}, \citenamefont {Jones},
  \citenamefont {Kaplan}, \citenamefont {Lam}, \citenamefont {Lazio},
  \citenamefont {Levin}, \citenamefont {Lommen}, \citenamefont {Lorimer},
  \citenamefont {Luo}, \citenamefont {Lynch}, \citenamefont {Madison},
  \citenamefont {Matthews}, \citenamefont {McLaughlin}, \citenamefont
  {McWilliams}, \citenamefont {Mingarelli}, \citenamefont {Ng}, \citenamefont
  {Nice}, \citenamefont {Pennucci}, \citenamefont {Ransom}, \citenamefont
  {Ray}, \citenamefont {Siemens}, \citenamefont {Simon}, \citenamefont
  {Spiewak}, \citenamefont {Stairs}, \citenamefont {Stinebring}, \citenamefont
  {Stovall}, \citenamefont {Swiggum}, \citenamefont {Taylor}, \citenamefont
  {Vallisneri}, \citenamefont {van Haasteren}, \citenamefont {Vigeland},\ and\
  \citenamefont {{and}}}]{arzoumanianNANOGrav11yearData2018}%
  \BibitemOpen
  \bibfield  {author} {\bibinfo {author} {\bibfnamefont {Z.}~\bibnamefont
  {Arzoumanian}}, \bibinfo {author} {\bibfnamefont {A.}~\bibnamefont
  {Brazier}}, \bibinfo {author} {\bibfnamefont {S.}~\bibnamefont
  {{Burke-Spolaor}}}, \emph {et~al.},\ }\bibfield  {title} {\bibinfo {title}
  {{\color{Gray}The {{NANOGrav}} 11-Year {{Data Set}}: {{High}}-Precision
  {{Timing}} of 45 {{Millisecond Pulsars}}}},\ }\href
  {https://doi.org/10.3847/1538-4365/aab5b0} {\bibfield  {journal} {\bibinfo
  {journal} {ApJS}\ }\textbf {\bibinfo {volume} {235}},\ \bibinfo {pages} {37}
  (\bibinfo {year} {2018})},\ \Eprint {https://arxiv.org/abs/1801.01837}
  {arXiv:1801.01837 [astro-ph.HE]} \BibitemShut {NoStop}%
\bibitem [{\citenamefont {Arzoumanian}\ \emph {et~al.}(2020)\citenamefont
  {Arzoumanian}, \citenamefont {Baker}, \citenamefont {Blumer}, \citenamefont
  {B{\'e}csy}, \citenamefont {Brazier}, \citenamefont {Brook}, \citenamefont
  {{Burke-Spolaor}}, \citenamefont {Chatterjee}, \citenamefont {Chen},
  \citenamefont {Cordes}, \citenamefont {Cornish}, \citenamefont {Crawford},
  \citenamefont {Cromartie}, \citenamefont {DeCesar}, \citenamefont {Demorest},
  \citenamefont {Dolch}, \citenamefont {Ellis}, \citenamefont {Ferrara},
  \citenamefont {Fiore}, \citenamefont {Fonseca}, \citenamefont
  {{Garver-Daniels}}, \citenamefont {Gentile}, \citenamefont {Good},
  \citenamefont {Hazboun}, \citenamefont {Holgado}, \citenamefont {Islo},
  \citenamefont {Jennings}, \citenamefont {Jones}, \citenamefont {Kaiser},
  \citenamefont {Kaplan}, \citenamefont {Kelley}, \citenamefont {Key},
  \citenamefont {Laal}, \citenamefont {Lam}, \citenamefont {Lazio},
  \citenamefont {Lorimer}, \citenamefont {Luo}, \citenamefont {Lynch},
  \citenamefont {Madison}, \citenamefont {McLaughlin}, \citenamefont
  {Mingarelli}, \citenamefont {Ng}, \citenamefont {Nice}, \citenamefont
  {Pennucci}, \citenamefont {Pol}, \citenamefont {Ransom}, \citenamefont {Ray},
  \citenamefont {{Shapiro-Albert}}, \citenamefont {Siemens}, \citenamefont
  {Simon}, \citenamefont {Spiewak}, \citenamefont {Stairs}, \citenamefont
  {Stinebring}, \citenamefont {Stovall}, \citenamefont {Sun}, \citenamefont
  {Swiggum}, \citenamefont {Taylor}, \citenamefont {Turner}, \citenamefont
  {Vallisneri}, \citenamefont {Vigeland},\ and\ \citenamefont
  {Witt}}]{arzoumanianNANOGrav125Yr2020}%
  \BibitemOpen
  \bibfield  {author} {\bibinfo {author} {\bibfnamefont {Z.}~\bibnamefont
  {Arzoumanian}}, \bibinfo {author} {\bibfnamefont {P.~T.}\ \bibnamefont
  {Baker}}, \bibinfo {author} {\bibfnamefont {H.}~\bibnamefont {Blumer}}, \emph
  {et~al.},\ }\bibfield  {title} {\bibinfo {title} {{\color{Gray}The
  {{NANOGrav}} 12.5 Yr {{Data Set}}: {{Search}} for an {{Isotropic Stochastic
  Gravitational}}-Wave {{Background}}}},\ }\href
  {https://doi.org/10.3847/2041-8213/abd401} {\bibfield  {journal} {\bibinfo
  {journal} {ApJL}\ }\textbf {\bibinfo {volume} {905}},\ \bibinfo {pages} {L34}
  (\bibinfo {year} {2020})},\ \Eprint {https://arxiv.org/abs/2009.04496}
  {arXiv:2009.04496 [astro-ph.HE]} \BibitemShut {NoStop}%
\bibitem [{\citenamefont {Larsen}\ \emph {et~al.}(2025)\citenamefont {Larsen},
  \citenamefont {Mingarelli}, \citenamefont {Baker}, \citenamefont {Hazboun},
  \citenamefont {Chen}, \citenamefont {Schult}, \citenamefont {Taylor},
  \citenamefont {Simon}, \citenamefont {Antoniadis} \emph
  {et~al.}}]{larsen2025lite}%
  \BibitemOpen
  \bibfield  {author} {\bibinfo {author} {\bibfnamefont {B.}~\bibnamefont
  {Larsen}}, \bibinfo {author} {\bibfnamefont {C.~M.~F.}\ \bibnamefont
  {Mingarelli}}, \bibinfo {author} {\bibfnamefont {P.~T.}\ \bibnamefont
  {Baker}}, \emph {et~al.},\ }\bibfield  {title} {\bibinfo {title}
  {{\color{Gray}Rapid construction of joint pulsar timing array data sets: the
  {Lite} method}},\ }\href {https://doi.org/10.1093/mnras/staf1420} {\bibfield
  {journal} {\bibinfo  {journal} {Monthly Notices of the Royal Astronomical
  Society}\ }\textbf {\bibinfo {volume} {542}},\ \bibinfo {pages} {3028}
  (\bibinfo {year} {2025})},\ \Eprint {https://arxiv.org/abs/2503.20949}
  {arXiv:2503.20949 [astro-ph.HE]} \BibitemShut {NoStop}%
\bibitem [{\citenamefont {Wright}\ \emph {et~al.}(2026)\citenamefont {Wright},
  \citenamefont {Wayt}, \citenamefont {Hazboun}, \citenamefont {Siemens},
  \citenamefont {van Haasteren}, \citenamefont {Schult},\ and\ \citenamefont
  {Taylor}}]{wright2026frankenstat}%
  \BibitemOpen
  \bibfield  {author} {\bibinfo {author} {\bibfnamefont {D.}~\bibnamefont
  {Wright}}, \bibinfo {author} {\bibfnamefont {K.}~\bibnamefont {Wayt}},
  \bibinfo {author} {\bibfnamefont {J.~S.}\ \bibnamefont {Hazboun}}, \emph
  {et~al.},\ }\bibfield  {title} {\bibinfo {title} {{\color{Gray}{FrankenStat}
  {I}: A new approach to pulsar timing array data combination}},\ }\href
  {https://doi.org/10.1103/h9rb-5vdm} {\bibfield  {journal} {\bibinfo
  {journal} {Physical Review Letters}\ }\textbf {\bibinfo {volume} {137}},\
  \bibinfo {pages} {041402} (\bibinfo {year} {2026})},\ \Eprint
  {https://arxiv.org/abs/2512.14807} {arXiv:2512.14807 [astro-ph.CO]}
  \BibitemShut {NoStop}%
\bibitem [{\citenamefont {Yu}\ and\ \citenamefont
  {Allen}(2025)}]{yuStochasticGravitationalwaveBackground2025}%
  \BibitemOpen
  \bibfield  {author} {\bibinfo {author} {\bibfnamefont {W.-W.}\ \bibnamefont
  {Yu}}\ and\ \bibinfo {author} {\bibfnamefont {B.}~\bibnamefont {Allen}},\
  }\href {https://arxiv.org/abs/2512.08666v3} {\bibinfo {title}
  {{\color{Gray}Stochastic Gravitational-Wave Background Search Using Data from
  Five Pulsar Timing Arrays}}} (\bibinfo {year} {2025}),\ \Eprint
  {https://arxiv.org/abs/2512.08666} {arXiv:2512.08666 [astro-ph.CO]}
  \BibitemShut {NoStop}%
\bibitem [{\citenamefont {Hobbs}\ \emph {et~al.}(2006)\citenamefont {Hobbs},
  \citenamefont {Edwards},\ and\ \citenamefont
  {Manchester}}]{hobbsTEMPO2NewPulsar2006}%
  \BibitemOpen
  \bibfield  {author} {\bibinfo {author} {\bibfnamefont {G.}~\bibnamefont
  {Hobbs}}, \bibinfo {author} {\bibfnamefont {R.}~\bibnamefont {Edwards}},\
  and\ \bibinfo {author} {\bibfnamefont {R.}~\bibnamefont {Manchester}},\
  }\bibfield  {title} {\bibinfo {title} {{\color{Gray}{TEMPO2}, a new
  pulsar-timing package -- {I}. {An} overview}},\ }\href
  {https://doi.org/10.1111/j.1365-2966.2006.10302.x} {\bibfield  {journal}
  {\bibinfo  {journal} {Monthly Notices of the Royal Astronomical Society}\
  }\textbf {\bibinfo {volume} {369}},\ \bibinfo {pages} {655} (\bibinfo {year}
  {2006})},\ \Eprint {https://arxiv.org/abs/astro-ph/0603381}
  {arXiv:astro-ph/0603381} \BibitemShut {NoStop}%
\bibitem [{\citenamefont {Lorimer}\ and\ \citenamefont
  {Kramer}(2004)}]{lorimerHandbookPulsarAstronomy2004}%
  \BibitemOpen
  \bibfield  {author} {\bibinfo {author} {\bibfnamefont {D.~R.}\ \bibnamefont
  {Lorimer}}\ and\ \bibinfo {author} {\bibfnamefont {M.}~\bibnamefont
  {Kramer}},\ }\href {https://ui.adsabs.harvard.edu/abs/2004hpa..book.....L}
  {\emph {\bibinfo {title} {Handbook of {{Pulsar Astronomy}}}}},\ Vol.~\bibinfo
  {volume} {4}\ (\bibinfo  {publisher} {Cambridge University Press},\ \bibinfo
  {year} {2004})\BibitemShut {NoStop}%
\bibitem [{\citenamefont {Damour}\ and\ \citenamefont
  {Deruelle}(1986)}]{damourGeneralRelativisticCelestial1986}%
  \BibitemOpen
  \bibfield  {author} {\bibinfo {author} {\bibfnamefont {T.}~\bibnamefont
  {Damour}}\ and\ \bibinfo {author} {\bibfnamefont {N.}~\bibnamefont
  {Deruelle}},\ }\bibfield  {title} {\bibinfo {title} {{\color{Gray}General
  relativistic celestial mechanics of binary systems. {II}. {The}
  post-newtonian timing formula}},\ }\href
  {https://www.numdam.org/item/AIHPA_1986__44_3_263_0/} {\bibfield  {journal}
  {\bibinfo  {journal} {Annales de l'I.H.P. Physique th\'eorique}\ }\textbf
  {\bibinfo {volume} {44}},\ \bibinfo {pages} {263} (\bibinfo {year}
  {1986})}\BibitemShut {NoStop}%
\bibitem [{\citenamefont {Edwards}\ \emph {et~al.}(2006)\citenamefont
  {Edwards}, \citenamefont {Hobbs},\ and\ \citenamefont
  {Manchester}}]{edwardsTempo2NewPulsar2006}%
  \BibitemOpen
  \bibfield  {author} {\bibinfo {author} {\bibfnamefont {R.~T.}\ \bibnamefont
  {Edwards}}, \bibinfo {author} {\bibfnamefont {G.~B.}\ \bibnamefont {Hobbs}},\
  and\ \bibinfo {author} {\bibfnamefont {R.~N.}\ \bibnamefont {Manchester}},\
  }\bibfield  {title} {\bibinfo {title} {{\color{Gray}{Tempo2}, a new pulsar
  timing package. {II}: {The} timing model and precision estimates}},\ }\href
  {https://doi.org/10.1111/j.1365-2966.2006.10870.x} {\bibfield  {journal}
  {\bibinfo  {journal} {Monthly Notices of the Royal Astronomical Society}\
  }\textbf {\bibinfo {volume} {372}},\ \bibinfo {pages} {1549} (\bibinfo {year}
  {2006})},\ \Eprint {https://arxiv.org/abs/astro-ph/0607664}
  {arXiv:astro-ph/0607664} \BibitemShut {NoStop}%
\bibitem [{\citenamefont {Luo}\ \emph {et~al.}(2021)\citenamefont {Luo},
  \citenamefont {Ransom}, \citenamefont {Demorest}, \citenamefont {Ray},
  \citenamefont {Archibald}, \citenamefont {Kerr}, \citenamefont {Jennings},
  \citenamefont {Bachetti}, \citenamefont {van Haasteren}, \citenamefont
  {Champagne}, \citenamefont {Colen}, \citenamefont {Phillips}, \citenamefont
  {Zimmerman}, \citenamefont {Stovall}, \citenamefont {Lam},\ and\
  \citenamefont {Jenet}}]{luoPINTModernSoftware2021}%
  \BibitemOpen
  \bibfield  {author} {\bibinfo {author} {\bibfnamefont {J.}~\bibnamefont
  {Luo}}, \bibinfo {author} {\bibfnamefont {S.}~\bibnamefont {Ransom}},
  \bibinfo {author} {\bibfnamefont {P.}~\bibnamefont {Demorest}}, \emph
  {et~al.},\ }\bibfield  {title} {\bibinfo {title} {{\color{Gray}{{PINT}}: {{A
  Modern Software Package}} for {{Pulsar Timing}}}},\ }\href
  {https://doi.org/10.3847/1538-4357/abe62f} {\bibfield  {journal} {\bibinfo
  {journal} {ApJ}\ }\textbf {\bibinfo {volume} {911}},\ \bibinfo {pages} {45}
  (\bibinfo {year} {2021})},\ \Eprint {https://arxiv.org/abs/2012.00074}
  {arXiv:2012.00074 [astro-ph.IM]} \BibitemShut {NoStop}%
\bibitem [{\citenamefont {Lentati}\ \emph {et~al.}(2014)\citenamefont
  {Lentati}, \citenamefont {Alexander}, \citenamefont {Hobson}, \citenamefont
  {Feroz}, \citenamefont {{van Haasteren}}, \citenamefont {Lee},\ and\
  \citenamefont {Shannon}}]{lentatiTemponestBayesianApproach2014}%
  \BibitemOpen
  \bibfield  {author} {\bibinfo {author} {\bibfnamefont {L.}~\bibnamefont
  {Lentati}}, \bibinfo {author} {\bibfnamefont {P.}~\bibnamefont {Alexander}},
  \bibinfo {author} {\bibfnamefont {M.~P.}\ \bibnamefont {Hobson}}, \emph
  {et~al.},\ }\bibfield  {title} {\bibinfo {title}
  {{\color{Gray}\texttt{TEMPONEST}: a {Bayesian} approach to pulsar timing
  analysis}},\ }\href {https://doi.org/10.1093/mnras/stt2122} {\bibfield
  {journal} {\bibinfo  {journal} {Monthly Notices of the Royal Astronomical
  Society}\ }\textbf {\bibinfo {volume} {437}},\ \bibinfo {pages} {3004}
  (\bibinfo {year} {2014})},\ \Eprint {https://arxiv.org/abs/1310.2120}
  {arXiv:1310.2120 [astro-ph.IM]} \BibitemShut {NoStop}%
\bibitem [{\citenamefont {Vigeland}\ and\ \citenamefont
  {Vallisneri}(2014)}]{vigelandBayesianInferencePulsartiming2014}%
  \BibitemOpen
  \bibfield  {author} {\bibinfo {author} {\bibfnamefont {S.~J.}\ \bibnamefont
  {Vigeland}}\ and\ \bibinfo {author} {\bibfnamefont {M.}~\bibnamefont
  {Vallisneri}},\ }\bibfield  {title} {\bibinfo {title}
  {{\color{Gray}{Bayesian} inference for pulsar-timing models}},\ }\href
  {https://doi.org/10.1093/mnras/stu312} {\bibfield  {journal} {\bibinfo
  {journal} {Mon Not R Astron Soc}\ }\textbf {\bibinfo {volume} {440}},\
  \bibinfo {pages} {1446} (\bibinfo {year} {2014})},\ \Eprint
  {https://arxiv.org/abs/1310.2606} {arXiv:1310.2606 [astro-ph.IM]}
  \BibitemShut {NoStop}%
\bibitem [{\citenamefont {Kaiser}\ \emph {et~al.}(2026)\citenamefont {Kaiser},
  \citenamefont {Hazboun}, \citenamefont {McLaughlin}, \citenamefont
  {Cromartie}, \citenamefont {Fonseca}, \citenamefont {Simon}, \citenamefont
  {Taylor}, \citenamefont {Vallisneri}, \citenamefont {Vigeland}, \citenamefont
  {Arzoumanian}, \citenamefont {Baker}, \citenamefont {Blumer}, \citenamefont
  {Brook}, \citenamefont {Cognard}, \citenamefont {DeCesar}, \citenamefont
  {Demorest}, \citenamefont {Dolch}, \citenamefont {Dong}, \citenamefont
  {Ellis}, \citenamefont {Ferdman}, \citenamefont {Ferrara}, \citenamefont
  {Fiore}, \citenamefont {{Garver-Daniels}}, \citenamefont {Gentile},
  \citenamefont {Good}, \citenamefont {Guillemot}, \citenamefont {Jennings},
  \citenamefont {Jones}, \citenamefont {Kaplan}, \citenamefont {Kerr},
  \citenamefont {Kirichenko}, \citenamefont {Lam}, \citenamefont {Lorimer},
  \citenamefont {Luo}, \citenamefont {Lynch}, \citenamefont {McEwen},
  \citenamefont {McKee}, \citenamefont {McMann}, \citenamefont {Meyers},
  \citenamefont {Naidu}, \citenamefont {Ng}, \citenamefont {Nice},
  \citenamefont {Parthasarathy}, \citenamefont {Pennucci}, \citenamefont
  {Perera}, \citenamefont {Pol}, \citenamefont {Radovan}, \citenamefont
  {Ransom}, \citenamefont {Ray}, \citenamefont {Spiewak}, \citenamefont
  {{Shapiro-Albert}}, \citenamefont {Stairs}, \citenamefont {Stovall},
  \citenamefont {Swiggum}, \citenamefont {Tan}, \citenamefont {Tendulkar},
  \citenamefont {Wahl},\ and\ \citenamefont
  {Zhu}}]{kaiserGeneralizedNonlinearBayesian2026a}%
  \BibitemOpen
  \bibfield  {author} {\bibinfo {author} {\bibfnamefont {A.~R.}\ \bibnamefont
  {Kaiser}}, \bibinfo {author} {\bibfnamefont {J.~S.}\ \bibnamefont {Hazboun}},
  \bibinfo {author} {\bibfnamefont {M.~A.}\ \bibnamefont {McLaughlin}}, \emph
  {et~al.},\ }\bibfield  {title} {\bibinfo {title} {{\color{Gray}Generalized
  {{Nonlinear Bayesian Pulsar Timing}} with {{ENTERPRISE}}}},\ }\href
  {https://doi.org/10.3847/1538-4357/ae99e3} {\bibfield  {journal} {\bibinfo
  {journal} {ApJ}\ }\textbf {\bibinfo {volume} {1010}},\ \bibinfo {pages} {4}
  (\bibinfo {year} {2026})},\ \Eprint {https://arxiv.org/abs/2608.18047}
  {arXiv:2608.18047 [astro-ph.HE]} \BibitemShut {NoStop}%
\bibitem [{\citenamefont
  {Susobhanan}(2025)}]{susobhananBayesianPulsarTiming2025b}%
  \BibitemOpen
  \bibfield  {author} {\bibinfo {author} {\bibfnamefont {A.}~\bibnamefont
  {Susobhanan}},\ }\bibfield  {title} {\bibinfo {title} {{\color{Gray}Bayesian
  {{Pulsar Timing}} and {{Noise Analysis}} with \texttt{{Vela}}.Jl: {{An
  Overview}}}},\ }\href {https://doi.org/10.3847/1538-4357/adaaec} {\bibfield
  {journal} {\bibinfo  {journal} {ApJ}\ }\textbf {\bibinfo {volume} {980}},\
  \bibinfo {pages} {165} (\bibinfo {year} {2025})},\ \Eprint
  {https://arxiv.org/abs/2412.15858} {arXiv:2412.15858 [astro-ph.IM]}
  \BibitemShut {NoStop}%
\bibitem [{\citenamefont {van Haasteren}\ and\ \citenamefont
  {Levin}(2013)}]{van2013understanding}%
  \BibitemOpen
  \bibfield  {author} {\bibinfo {author} {\bibfnamefont {R.}~\bibnamefont {van
  Haasteren}}\ and\ \bibinfo {author} {\bibfnamefont {Y.}~\bibnamefont
  {Levin}},\ }\bibfield  {title} {\bibinfo {title} {{\color{Gray}Understanding
  and analysing time-correlated stochastic signals in pulsar timing}},\ }\href
  {https://doi.org/10.1093/mnras/sts097} {\bibfield  {journal} {\bibinfo
  {journal} {Monthly Notices of the Royal Astronomical Society}\ }\textbf
  {\bibinfo {volume} {428}},\ \bibinfo {pages} {1147} (\bibinfo {year}
  {2013})},\ \Eprint {https://arxiv.org/abs/1202.5932} {arXiv:1202.5932
  [astro-ph.IM]} \BibitemShut {NoStop}%
\bibitem [{\citenamefont {{van Haasteren}}\ and\ \citenamefont
  {Vallisneri}(2014)}]{vanhaasterenNewAdvancesGaussianprocess2014}%
  \BibitemOpen
  \bibfield  {author} {\bibinfo {author} {\bibfnamefont {R.}~\bibnamefont {{van
  Haasteren}}}\ and\ \bibinfo {author} {\bibfnamefont {M.}~\bibnamefont
  {Vallisneri}},\ }\bibfield  {title} {\bibinfo {title} {{\color{Gray}New
  advances in the {{Gaussian}}-process approach to pulsar-timing data
  analysis}},\ }\href {https://doi.org/10.1103/PhysRevD.90.104012} {\bibfield
  {journal} {\bibinfo  {journal} {Phys. Rev. D}\ }\textbf {\bibinfo {volume}
  {90}},\ \bibinfo {pages} {104012} (\bibinfo {year} {2014})},\ \Eprint
  {https://arxiv.org/abs/1407.1838} {arXiv:1407.1838 [gr-qc]} \BibitemShut
  {NoStop}%
\bibitem [{\citenamefont {Ding}\ \emph {et~al.}(2023)\citenamefont {Ding},
  \citenamefont {Deller}, \citenamefont {Stappers}, \citenamefont {Lazio},
  \citenamefont {Kaplan}, \citenamefont {Chatterjee}, \citenamefont {Brisken},
  \citenamefont {Cordes}, \citenamefont {Freire}, \citenamefont {Fonseca} \emph
  {et~al.}}]{ding2023mspsrpi}%
  \BibitemOpen
  \bibfield  {author} {\bibinfo {author} {\bibfnamefont {H.}~\bibnamefont
  {Ding}}, \bibinfo {author} {\bibfnamefont {A.}~\bibnamefont {Deller}},
  \bibinfo {author} {\bibfnamefont {B.}~\bibnamefont {Stappers}}, \emph
  {et~al.},\ }\bibfield  {title} {\bibinfo {title} {{\color{Gray}The
  {MSPSR}$\pi$ catalogue: {VLBA} astrometry of 18 millisecond pulsars}},\
  }\href {https://doi.org/10.1093/mnras/stac3725} {\bibfield  {journal}
  {\bibinfo  {journal} {Monthly Notices of the Royal Astronomical Society}\
  }\textbf {\bibinfo {volume} {519}},\ \bibinfo {pages} {4982} (\bibinfo {year}
  {2023})},\ \Eprint {https://arxiv.org/abs/2212.06351} {arXiv:2212.06351
  [astro-ph.HE]} \BibitemShut {NoStop}%
\bibitem [{\citenamefont {Agazie}\ \emph
  {et~al.}(2023{\natexlab{b}})\citenamefont {Agazie}, \citenamefont {Alam},
  \citenamefont {Anumarlapudi}, \citenamefont {Archibald}, \citenamefont
  {Arzoumanian}, \citenamefont {Baker}, \citenamefont {Blecha}, \citenamefont
  {Bonidie}, \citenamefont {Brazier}, \citenamefont {Brook}, \citenamefont
  {{Burke-Spolaor}}, \citenamefont {B{\'e}csy}, \citenamefont {Chapman},
  \citenamefont {Charisi}, \citenamefont {Chatterjee}, \citenamefont {Cohen},
  \citenamefont {Cordes}, \citenamefont {Cornish}, \citenamefont {Crawford},
  \citenamefont {Cromartie}, \citenamefont {Crowter}, \citenamefont {DeCesar},
  \citenamefont {Demorest}, \citenamefont {Dolch}, \citenamefont {Drachler},
  \citenamefont {Ferrara}, \citenamefont {Fiore}, \citenamefont {Fonseca},
  \citenamefont {Freedman}, \citenamefont {{Garver-Daniels}}, \citenamefont
  {Gentile}, \citenamefont {Glaser}, \citenamefont {Good}, \citenamefont
  {G{\"u}ltekin}, \citenamefont {Hazboun}, \citenamefont {Jennings},
  \citenamefont {Jessup}, \citenamefont {Johnson}, \citenamefont {Jones},
  \citenamefont {Kaiser}, \citenamefont {Kaplan}, \citenamefont {Kelley},
  \citenamefont {Kerr}, \citenamefont {Key}, \citenamefont {Kuske},
  \citenamefont {Laal}, \citenamefont {Lam}, \citenamefont {Lamb},
  \citenamefont {Lazio}, \citenamefont {Lewandowska}, \citenamefont {Lin},
  \citenamefont {Liu}, \citenamefont {Lorimer}, \citenamefont {Luo},
  \citenamefont {Lynch}, \citenamefont {Ma}, \citenamefont {Madison},
  \citenamefont {Maraccini}, \citenamefont {McEwen}, \citenamefont {McKee},
  \citenamefont {McLaughlin}, \citenamefont {McMann}, \citenamefont {Meyers},
  \citenamefont {Mingarelli}, \citenamefont {Mitridate}, \citenamefont {Ng},
  \citenamefont {Nice}, \citenamefont {Ocker}, \citenamefont {Olum},
  \citenamefont {Panciu}, \citenamefont {Pennucci}, \citenamefont {Perera},
  \citenamefont {Pol}, \citenamefont {Radovan}, \citenamefont {Ransom},
  \citenamefont {Ray}, \citenamefont {Romano}, \citenamefont {Salo},
  \citenamefont {Sardesai}, \citenamefont {Schmiedekamp}, \citenamefont
  {Schmiedekamp}, \citenamefont {Schmitz}, \citenamefont {{Shapiro-Albert}},
  \citenamefont {Siemens}, \citenamefont {Simon}, \citenamefont {Siwek},
  \citenamefont {Stairs}, \citenamefont {Stinebring}, \citenamefont {Stovall},
  \citenamefont {Susobhanan}, \citenamefont {Swiggum}, \citenamefont {Taylor},
  \citenamefont {Turner}, \citenamefont {Unal}, \citenamefont {Vallisneri},
  \citenamefont {Vigeland}, \citenamefont {Wahl}, \citenamefont {Wang},
  \citenamefont {Witt}, \citenamefont {Young},\ and\ \citenamefont
  {Collaboration}}]{agazieNANOGrav15Yr2023}%
  \BibitemOpen
  \bibfield  {author} {\bibinfo {author} {\bibfnamefont {G.}~\bibnamefont
  {Agazie}}, \bibinfo {author} {\bibfnamefont {M.~F.}\ \bibnamefont {Alam}},
  \bibinfo {author} {\bibfnamefont {A.}~\bibnamefont {Anumarlapudi}}, \emph
  {et~al.},\ }\bibfield  {title} {\bibinfo {title} {{\color{Gray}The
  {{NANOGrav}} 15 Yr {{Data Set}}: {{Observations}} and {{Timing}} of 68
  {{Millisecond Pulsars}}}},\ }\href {https://doi.org/10.3847/2041-8213/acda9a}
  {\bibfield  {journal} {\bibinfo  {journal} {ApJL}\ }\textbf {\bibinfo
  {volume} {951}},\ \bibinfo {pages} {L9} (\bibinfo {year}
  {2023}{\natexlab{b}})},\ \Eprint {https://arxiv.org/abs/2306.16217}
  {arXiv:2306.16217 [astro-ph.HE]} \BibitemShut {NoStop}%
\bibitem [{\citenamefont {van Haasteren}(2026)}]{nltiming}%
  \BibitemOpen
  \bibfield  {author} {\bibinfo {author} {\bibfnamefont {R.}~\bibnamefont {van
  Haasteren}},\ }\href@noop {} {\bibinfo {title} {{\color{Gray}{nltiming}:
  nonlinear timing-model inference for pulsar timing}}},\ \bibinfo
  {howpublished} {\url{https://github.com/vhaasteren/nltiming}} (\bibinfo
  {year} {2026})\BibitemShut {NoStop}%
\bibitem [{\citenamefont {Ellis}\ \emph {et~al.}(2020)\citenamefont {Ellis},
  \citenamefont {Vallisneri}, \citenamefont {Taylor},\ and\ \citenamefont
  {Baker}}]{ellisENTERPRISEEnhancedNumerical2020}%
  \BibitemOpen
  \bibfield  {author} {\bibinfo {author} {\bibfnamefont {J.~A.}\ \bibnamefont
  {Ellis}}, \bibinfo {author} {\bibfnamefont {M.}~\bibnamefont {Vallisneri}},
  \bibinfo {author} {\bibfnamefont {S.~R.}\ \bibnamefont {Taylor}},\ and\
  \bibinfo {author} {\bibfnamefont {P.~T.}\ \bibnamefont {Baker}},\ }\href
  {https://doi.org/10.5281/zenodo.4059815} {\bibinfo {title}
  {{\color{Gray}{{ENTERPRISE}}: {{Enhanced Numerical Toolbox Enabling}} a
  {{Robust PulsaR Inference SuitE}}}}},\ \bibinfo {howpublished} {Zenodo}
  (\bibinfo {year} {2020})\BibitemShut {NoStop}%
\bibitem [{\citenamefont {Vallisneri}\ \emph {et~al.}(2025)\citenamefont
  {Vallisneri} \emph {et~al.}}]{discovery}%
  \BibitemOpen
  \bibfield  {author} {\bibinfo {author} {\bibfnamefont {M.}~\bibnamefont
  {Vallisneri}} \emph {et~al.},\ }\href@noop {} {\bibinfo {title}
  {{\color{Gray}{discovery}: {JAX}-based pulsar timing array analysis}}},\
  \bibinfo {howpublished} {\url{https://github.com/nanograv/discovery}}
  (\bibinfo {year} {2025})\BibitemShut {NoStop}%
\bibitem [{\citenamefont {Crisostomi}\ \emph {et~al.}(2025)\citenamefont
  {Crisostomi}, \citenamefont {van Haasteren}, \citenamefont {Meyers},\ and\
  \citenamefont {Vallisneri}}]{crisostomiDiagonalApproximationsImproved2025}%
  \BibitemOpen
  \bibfield  {author} {\bibinfo {author} {\bibfnamefont {M.}~\bibnamefont
  {Crisostomi}}, \bibinfo {author} {\bibfnamefont {R.}~\bibnamefont {van
  Haasteren}}, \bibinfo {author} {\bibfnamefont {P.~M.}\ \bibnamefont
  {Meyers}},\ and\ \bibinfo {author} {\bibfnamefont {M.}~\bibnamefont
  {Vallisneri}},\ }\href {https://doi.org/10.48550/arXiv.2506.13866} {\bibinfo
  {title} {{\color{Gray}{Beyond} diagonal approximations: improved covariance
  modeling for pulsar timing array data analysis}}} (\bibinfo {year} {2025}),\
  \Eprint {https://arxiv.org/abs/2506.13866} {arXiv:2506.13866 [astro-ph]}
  \BibitemShut {NoStop}%
\bibitem [{\citenamefont {{Cardinal Trembley}}\ \emph
  {et~al.}(2026)\citenamefont {{Cardinal Trembley}}, \citenamefont
  {Goncharov},\ and\ \citenamefont {van
  Haasteren}}]{jacobTensionStatistic2026}%
  \BibitemOpen
  \bibfield  {author} {\bibinfo {author} {\bibnamefont {{Cardinal Trembley}}},
  \bibinfo {author} {\bibfnamefont {B.}~\bibnamefont {Goncharov}},\ and\
  \bibinfo {author} {\bibfnamefont {R.}~\bibnamefont {van Haasteren}},\
  }\bibfield  {title} {\bibinfo {title} {{\color{Gray}Tension statistics for
  pulsar-timing-array constraints on continuous gravitational-wave sources}}}
  (\bibinfo {year} {2026}),\ \bibinfo {note} {in preparation; placeholder
  entry, fix author list and title}\BibitemShut {NoStop}%
\bibitem [{\citenamefont {Raveri}\ and\ \citenamefont
  {Hu}(2019)}]{raveriConcordanceDiscordance2019}%
  \BibitemOpen
  \bibfield  {author} {\bibinfo {author} {\bibfnamefont {M.}~\bibnamefont
  {Raveri}}\ and\ \bibinfo {author} {\bibfnamefont {W.}~\bibnamefont {Hu}},\
  }\bibfield  {title} {\bibinfo {title} {{\color{Gray}Concordance and
  discordance in cosmology}},\ }\href
  {https://doi.org/10.1103/PhysRevD.99.043506} {\bibfield  {journal} {\bibinfo
  {journal} {Physical Review D}\ }\textbf {\bibinfo {volume} {99}},\ \bibinfo
  {pages} {043506} (\bibinfo {year} {2019})},\ \Eprint
  {https://arxiv.org/abs/1806.04649} {arXiv:1806.04649 [astro-ph.CO]}
  \BibitemShut {NoStop}%
\bibitem [{\citenamefont {Hoffman}\ \emph {et~al.}(2014)\citenamefont
  {Hoffman}, \citenamefont {Gelman} \emph {et~al.}}]{hoffman2014no}%
  \BibitemOpen
  \bibfield  {author} {\bibinfo {author} {\bibfnamefont {M.~D.}\ \bibnamefont
  {Hoffman}}, \bibinfo {author} {\bibfnamefont {A.}~\bibnamefont {Gelman}},
  \emph {et~al.},\ }\bibfield  {title} {\bibinfo {title} {{\color{Gray}The
  No-U-Turn {Sampler}: {Adaptively} {Setting} {Path} {Lengths} in {Hamiltonian}
  {Monte} {Carlo}.}},\ }\href {http://jmlr.org/papers/v15/hoffman14a.html}
  {\bibfield  {journal} {\bibinfo  {journal} {J. Mach. Learn. Res.}\ }\textbf
  {\bibinfo {volume} {15}},\ \bibinfo {pages} {1593} (\bibinfo {year}
  {2014})},\ \Eprint {https://arxiv.org/abs/1111.4246} {arXiv:1111.4246
  [stat.CO]} \BibitemShut {NoStop}%
\bibitem [{\citenamefont {Freire}\ and\ \citenamefont
  {Wex}(2010)}]{freireOrthometricParametrizationShapiro2010}%
  \BibitemOpen
  \bibfield  {author} {\bibinfo {author} {\bibfnamefont {P.~C.~C.}\
  \bibnamefont {Freire}}\ and\ \bibinfo {author} {\bibfnamefont
  {N.}~\bibnamefont {Wex}},\ }\bibfield  {title} {\bibinfo {title}
  {{\color{Gray}The orthometric parametrization of the {{Shapiro}} delay and an
  improved test of general relativity with binary pulsars}},\ }\href
  {https://doi.org/10.1111/j.1365-2966.2010.17319.x} {\bibfield  {journal}
  {\bibinfo  {journal} {Monthly Notices of the Royal Astronomical Society}\
  }\textbf {\bibinfo {volume} {409}},\ \bibinfo {pages} {199} (\bibinfo {year}
  {2010})}\BibitemShut {NoStop}%
\bibitem [{\citenamefont {Blandford}\ and\ \citenamefont
  {Teukolsky}(1976)}]{blandfordArrivaltimeAnalysisPulsar1976}%
  \BibitemOpen
  \bibfield  {author} {\bibinfo {author} {\bibfnamefont {R.}~\bibnamefont
  {Blandford}}\ and\ \bibinfo {author} {\bibfnamefont {S.~A.}\ \bibnamefont
  {Teukolsky}},\ }\bibfield  {title} {\bibinfo {title}
  {{\color{Gray}Arrival-time analysis for a pulsar in a binary system.}},\
  }\href {https://doi.org/10.1086/154315} {\bibfield  {journal} {\bibinfo
  {journal} {The Astrophysical Journal}\ }\textbf {\bibinfo {volume} {205}},\
  \bibinfo {pages} {580} (\bibinfo {year} {1976})}\BibitemShut {NoStop}%
\bibitem [{\citenamefont {Damour}\ and\ \citenamefont
  {Deruelle}(1985)}]{damourGeneralRelativisticCelestial1985}%
  \BibitemOpen
  \bibfield  {author} {\bibinfo {author} {\bibfnamefont {T.}~\bibnamefont
  {Damour}}\ and\ \bibinfo {author} {\bibfnamefont {N.}~\bibnamefont
  {Deruelle}},\ }\bibfield  {title} {\bibinfo {title} {{\color{Gray}General
  relativistic celestial mechanics of binary systems. {I}. {T}he post-newtonian
  motion}},\ }\href {https://www.numdam.org/item/AIHPA_1985__43_1_107_0/}
  {\bibfield  {journal} {\bibinfo  {journal} {Annales de l'I.H.P. Physique
  th\'eorique}\ }\textbf {\bibinfo {volume} {43}},\ \bibinfo {pages} {107}
  (\bibinfo {year} {1985})}\BibitemShut {NoStop}%
\bibitem [{\citenamefont {Lange}\ \emph {et~al.}(2001)\citenamefont {Lange},
  \citenamefont {Camilo}, \citenamefont {Wex}, \citenamefont {Kramer},
  \citenamefont {Backer}, \citenamefont {Lyne},\ and\ \citenamefont
  {Doroshenko}}]{langePrecisionTimingMeasurements2001}%
  \BibitemOpen
  \bibfield  {author} {\bibinfo {author} {\bibfnamefont {{\relax
  Ch}.}~\bibnamefont {Lange}}, \bibinfo {author} {\bibfnamefont
  {F.}~\bibnamefont {Camilo}}, \bibinfo {author} {\bibfnamefont
  {N.}~\bibnamefont {Wex}}, \emph {et~al.},\ }\bibfield  {title} {\bibinfo
  {title} {{\color{Gray}Precision Timing Measurements of {{PSR J1012}}+5307}},\
  }\href {https://doi.org/10.1046/j.1365-8711.2001.04606.x} {\bibfield
  {journal} {\bibinfo  {journal} {Mon Not R Astron Soc}\ }\textbf {\bibinfo
  {volume} {326}},\ \bibinfo {pages} {274} (\bibinfo {year} {2001})},\ \Eprint
  {https://arxiv.org/abs/astro-ph/0102309} {arXiv:astro-ph/0102309}
  \BibitemShut {NoStop}%
\bibitem [{\citenamefont
  {Kopeikin}(1995)}]{kopeikinPossibleImplicationsOrbital1995}%
  \BibitemOpen
  \bibfield  {author} {\bibinfo {author} {\bibfnamefont {S.~M.}\ \bibnamefont
  {Kopeikin}},\ }\bibfield  {title} {\bibinfo {title} {{\color{Gray}On
  {Possible} {Implications} of {Orbital} {Parallaxes} of {Wide} {Orbit}
  {Binary} {Pulsars} and {Their} {Measurability}}},\ }\href
  {https://doi.org/10.1086/187731} {\bibfield  {journal} {\bibinfo  {journal}
  {The Astrophysical Journal Letters}\ }\textbf {\bibinfo {volume} {439}},\
  \bibinfo {pages} {L5} (\bibinfo {year} {1995})}\BibitemShut {NoStop}%
\bibitem [{\citenamefont {Kopeikin}(1996)}]{kopeikinProperMotionBinary1996}%
  \BibitemOpen
  \bibfield  {author} {\bibinfo {author} {\bibfnamefont {S.~M.}\ \bibnamefont
  {Kopeikin}},\ }\bibfield  {title} {\bibinfo {title} {{\color{Gray}{Proper}
  {Motion} of {Binary} {Pulsars} as a {Source} of {Secular} {Variations} of
  {Orbital} {Parameters}}},\ }\href {https://doi.org/10.1086/310201} {\bibfield
   {journal} {\bibinfo  {journal} {The Astrophysical Journal Letters}\ }\textbf
  {\bibinfo {volume} {467}},\ \bibinfo {pages} {L93} (\bibinfo {year}
  {1996})}\BibitemShut {NoStop}%
\bibitem [{\citenamefont {Damour}\ and\ \citenamefont
  {Taylor}(1992)}]{damourStrongFieldTests1992}%
  \BibitemOpen
  \bibfield  {author} {\bibinfo {author} {\bibfnamefont {T.}~\bibnamefont
  {Damour}}\ and\ \bibinfo {author} {\bibfnamefont {J.~H.}\ \bibnamefont
  {Taylor}},\ }\bibfield  {title} {\bibinfo {title} {{\color{Gray}Strong-field
  tests of relativistic gravity and binary pulsars}},\ }\href
  {https://doi.org/10.1103/PhysRevD.45.1840} {\bibfield  {journal} {\bibinfo
  {journal} {Physical Review D}\ }\textbf {\bibinfo {volume} {45}},\ \bibinfo
  {pages} {1840} (\bibinfo {year} {1992})}\BibitemShut {NoStop}%
\bibitem [{\citenamefont {Zhu}\ \emph {et~al.}(2019)\citenamefont {Zhu},
  \citenamefont {Desvignes}, \citenamefont {Wex} \emph
  {et~al.}}]{zhuTestsGravitationalSymmetries2019}%
  \BibitemOpen
  \bibfield  {author} {\bibinfo {author} {\bibfnamefont {W.~W.}\ \bibnamefont
  {Zhu}}, \bibinfo {author} {\bibfnamefont {G.}~\bibnamefont {Desvignes}},
  \bibinfo {author} {\bibfnamefont {N.}~\bibnamefont {Wex}}, \emph {et~al.},\
  }\bibfield  {title} {\bibinfo {title} {{\color{Gray}Tests of gravitational
  symmetries with pulsar binary {J1713+0747}}},\ }\href
  {https://doi.org/10.1093/mnras/sty2905} {\bibfield  {journal} {\bibinfo
  {journal} {Monthly Notices of the Royal Astronomical Society}\ }\textbf
  {\bibinfo {volume} {482}},\ \bibinfo {pages} {3249} (\bibinfo {year}
  {2019})},\ \Eprint {https://arxiv.org/abs/1802.09206} {arXiv:1802.09206}
  \BibitemShut {NoStop}%
\end{thebibliography}%

\end{document}